\documentclass[twocolumn,preprintnumbers]{revtex4}

\usepackage{graphicx}
\usepackage{epsf}
\usepackage{amsmath,amssymb}
\newcommand{\bvec}{\boldsymbol}

\begin{document}
%\preprint{KUNS-2503}
\title{Toroidal, compressive, and $E1$ properties of low-energy dipole modes in $^{10}$Be}
\author{Yoshiko Kanada-En'yo}
\affiliation{Department of Physics, Kyoto University, Kyoto 606-8502, Japan}
\author{Yuki Shikata}
\affiliation{Department of Physics, Kyoto University, Kyoto 606-8502, Japan}
\begin{abstract}
We studied dipole excitations in $^{10}$Be based on an extended version of the antisymmetrized
molecular dynamics, which can describe 1p-1h excitations and 
large amplitude cluster modes. 
Toroidal and compressive dipole operators are found to be good proves to separate 
the low-energy and high-energy parts of the isoscalar dipole excitations, respectively.
Two low-energy $1^-$ states, 
the toroidal dominant $1^-_1$ state at $E\sim 8$ MeV and the $E1$ dominant 
$1^-_2$ state at $E\sim 16$ MeV, were obtained.
By analysis of transition current densities, 
the $1^-_1$ states is understood as a toroidal dipole mode with
exotic toroidal neutron flow caused by rotation of a deformed $^6\textrm{He}$ cluster,
whereas the  $1^-_2$ state is regarded as 
a neutron-skin oscillation mode, which are characterized by 
surface neutron flow with inner isoscalar flow caused 
by the surface neutron oscillation against the $2\alpha$ core.

\end{abstract}
\maketitle
\section{Introduction}
In recent development in physics of unstable nuclei, 
low-energy dipole excitations have being attracting great interests 
and intensively studied in experimental and theoretical works (see, e.g.,  reviews in Refs.~\cite{Paar:2007bk,aumann-rev,Savran:2013bha,Bracco:2015hca} and references therein). 
Remarkable progress has been made in these years in experimentally studying isospin characters of low-energy dipole excitations 
for various nuclei, in particular, neutron-rich nuclei \cite{Bracco:2015hca,Endres:2010zw, Derya:2014yqk,Crespi:2014tka,Crespi:2015sfa,Nakatsuka:2017dhs}.
%\cite{kobayashi89,Ieki:1992mc,Sackett:1993zz,Shimoura:1994me,Zinser:1997da,Aumann:1999mb,Nakamura:2006zz,
%Kanungo:2015dna,Millener:1983zz,
%Hansen:1987mc,Bertsch:1990zza,Suzuki:1990uq,Honma90,Bertsch:1991zz,Sagawa92,Csoto:1994ji,Suzuki00,Garrido:2002ws,Myo:2003bh,
%chulkov,Bertulani:2007rm,Hagino:2007rn,Hagino:2009sj,Baye:2009zz,Kikuchi:2010zzb,Pinilla:2012zz,Kikuchi:2013ula,Nakamura,Palit,
%Fukuda:2004ct,Myo98,Sagawa:2001kx,
%Tohyama95,Hamamoto96a,Hamamoto96b,Sagawa:1999zz,Colo:2001fz,Nakatsukasa:2004ys,KanadaEn'yo:2005wd,
%VanIsacker:1992zz,Catara,Matsuo:2001wy,Vretenar:2001hs,Goriely:2002cx,Paar:2002gz,Tsoneva:2003gv,Matsuo:2004pr,
%Piekarewicz:2006ip,Terasaki:2006ts,Liang:2007ri,Tsoneva:2007fk,Paar:2007bk,Yoshida:2008rw,Co':2009gi,Martini:2011gy,Inakura:2011mv,
%RocaMaza:2011ug,Ebata:2014aaa,Bacca:2014rta,Piekarewicz:2010fa,Carbone:2010az,Reinhard:2012vw,Inakura:2013waa,
%Govaert:1998zz,Herzberg:1999tb,Leistenschneider:2001zz,Ryezayeva:2002zz,Tryggestad:2003gz,Hartmann:2004zz,
%Adrich:2005zz,Gibelin:2008zz,Klimkiewicz:2007zz,Schwengner:2008rk,Wieland:2009zz,Endres:2009zz,Endres:2010zw,
%Tamii:2011pv}.
For stable nuclei, familiar dipole excitations in high-energy region known to be
giant dipole resonances (GDRs) 
have been systematically observed in various nuclei by means of photonuclear reactions
and $\alpha$ (or $^6$Li) inelastic scatterings, which can probe isovector (IV) and isoscalar (IS) 
dipole excitations, respectively 
\cite{Berman:1975tt, morsch80,Harakeh:1981zz,giai81,Colo:2003zm}.
The low-energy dipole strengths below the GDR energy are often called as pigmy dipole resonances (PDR) 
and considered to be new excitation modes decoupled from the GDR modes. 
In  IV dipole ($E1$) excitations, 
the IV GDR (IVGDR) is understood as the
collective vibration mode originating in the opposite oscillation between protons and neutrons.
To understand the low-energy $E1$ strengths, 
a picture of surface neutron oscillation against a core 
has been proposed. In this paper, we call this mode ``neutron-skin oscillation mode''. (In some works, the word ``PDR'' is 
used to call this specific mode.)   
Also in IS dipole (ISD) excitations, low-energy strengths known 
in such nuclei as $^{16}$O, $^{40}$Ca, and $^{208}$Pb\cite{Harakeh:1981zz,Decowski:1981pcz,Poelhekken:1992gvp}
have been discussed in relation to 
an exotic dipole mode, i.e., the toroidal dipole (TD) mode 
\cite{Paar:2007bk,Colo:2003zm,Colo:2000br,Vretenar:2001te,Ryezayeva:2002zz,Papakonstantinou:2010ja,Kvasil:2011yk,Repko:2012rj,Nesterenko:2016qiw}

The TD mode carries vorticity and its character is much 
different from the compressive dipole (CD) mode, which is the
normal mode for the ISGDR.
In this decade, IS and IV properties of low-energy dipole excitations 
have been intensively studied to clarify essential nature of low-energy dipole modes \cite{Paar:2007bk,Savran:2013bha,Bracco:2015hca}.
One of the interesting problems is whether the vorticity origin TD mode arises as low-energy resonances in nuclear systems. 
In works with quasiparticle phonon model (QPM) and random phase approximation (RPA) for
such nuclei as $^{208}\textrm{Pb}$ and $^{132}\textrm{Sn}$
\cite{Ryezayeva:2002zz,Kvasil:2011yk,Repko:2012rj,Nesterenko:2016qiw}, 
it has been shown that 
the TD mode dominates the low-energy part of the ISD strengths whereas the CD mode 
mainly excites the high-energy part for the ISGDR, indicating that the toroidal property is a key 
for low-energy dipole resonances.
The TD dominant nature of the low-energy $E1$ resonances has been demonstrated 
by toroidal flow in transition current densities
\cite{Paar:2007bk,Vretenar:2001te,Ryezayeva:2002zz,Kvasil:2011yk,Repko:2012rj,Nesterenko:2016qiw}.
In the works of Refs.~\cite{Repko:2012rj,Nesterenko:2016qiw}, no low-energy $E1$ resonances for the pure neutron-skin oscillation mode has been obtained.

For light stable nuclei, low-energy IS strengths can be good probes also for
cluster states \cite{Suzuki:1989zza,Kawabata:2005ta,
Funaki:2006gt,KanadaEn'yo:2006bd,Wakasa:2007zza,Yamada:2011ri,Ichikawa:2011ji,kawabata13,Kanada-En'yo:2013dma,Kanada-Enyo:2015vwc,Chiba:2015zxa,Chiba:2015khu,Chiba:2016zyz}.
As discussed by Yamada {\it et al.} \cite{Yamada:2011ri} and Chiba {\it et al.} \cite{Chiba:2015khu}, 
cluster states can be strongly excited by IS compressive modes such as IS monopole (ISM) and ISD modes.
For instance,  in $^{12}$C and $^{16}$O, 
the enhanced low-energy ISM strengths are understood as cluster states. 
Moreover, ignificant low-energy ISD strengths observed in such nuclei as $^{12}$C and $^{16}$O 
\cite{John:2003ke,Lui:2001xh} are considered to probe $1^-$ cluster states
as discussed in Ref.~\cite{Kanada-Enyo:2015vwc}.

In light neutron-rich nuclei, a further rich variety of cluster states are expected to appear 
in excited states because of excess neutrons surrounding clusters (see, for example, Refs.~\cite{Oertzen-rev,freer07-rev,KanadaEn'yo:2012bj}
and references therein).  An typical example is the cluster structures consisting of a $2\alpha$-cluster core and 
surrounding valence neutrons in neutron-rich Be (see also a review in Ref.~\cite{Ito2014-rev}).
The ISM strengths in Be isotopes have been theoretically studied by cluster models 
and suggested to be a good probe for cluster states \cite{Ito2014-rev,Ito:2011zza,Kanada-Enyo:2016jnq}.
One of the authors, Y. K-E., has studied  
the $E1$ and compressive ISD strengths of Be isotopes and discussed the dipole 
excitations for cluster states \cite{Kanada-Enyo:2015knx}. 

Our main aim is to investigate toroidal nature of the low-energy dipole excitations in 
$^{10}$Be. 
We are going to show how the toroidal, compressive, and $E1$ operators excite
low-energy cluster states and high-energy GDRs.
A particular attention is paid on two components, the toroidal and the neutron-skin oscillation modes, 
in the low-energy dipole strengths for cluster states.

Usually, either of a mean-field approach or a cluster model fails to 
describe low-energy cluster states and high-energy GRs in a unified manner
because cluster states are large amplitude modes of highly correlated many nucleons
beyond mean-field approaches, whereas GR modes are collective vibrations 
described by coherent 1p-1h excitations,  which are not contained in ordinary cluster model space.
To take into account large amplitude cluster modes and coherent 1p-1h excitations, we have recently 
developed a new method based on the antisymmetrized molecular dynamics (AMD) \cite{Ono:1991uz,Ono:1992uy,KanadaEnyo:1995tb,KanadaEnyo:1995ir,KanadaEn'yo:2001qw,KanadaEn'yo:2012bj}:
the shifted-basis AMD (sAMD) combined with the cluster  generator coordinate method (GCM) 
\cite{Kanada-En'yo:2013dma,Kanada-Enyo:2015knx,Kanada-Enyo:2015vwc}.
In the method, we superpose various configurations including 1p-1h and cluster states expressed by
 AMD wave functions. In the framework, angular-momentum and parity projections are microscopically 
performed and the center-of-mass motion is exactly removed.
The method has been applied to investigate ISM excitations in $^{16}$O and ISM and ISD in $^{12}$C, and proved to be a useful method 
to describe mopole and dipole excitations in a wide energy region including 
low-energy cluster modes and higher-energy GR modes in a unified framework. 
In our previous work \cite{Kanada-Enyo:2015knx}, we  applied the method for 
$E1$ and ISD excitations in neutron-rich Be isotopes and showed
that low-energy $E1$ and ISD strengths for cluster states 
appear separating from high-energy strengths for GDRs.
In this paper, we investigate toroidal, compressive, and $E1$ properties 
of dipole excitations in $^{10}$Be based on reanalysis of the previous calculation.
By analysis of  transition current densities in the low-energy dipole excitations, 
we show a toroidal feature of cluster states.
We also perform a cluster model analysis to obtain intuitive understanding of 
toroidal dominance in the $1^-_1$ state and $E1$ dominance in the $1^-_2$ states.

This paper is organized as follows. 
The definition of dipole operators and transitions are explained in Sec.~\ref{sec:dipole}.
The calculation scheme and results of dipole excitations in $^{10}$Be are shown in Sec.~\ref{sec:10Be}, 
and properties of low-energy dipole modes are discussed in Sec.~\ref{sec:discussions}.
The paper concludes with a summary and an outlook in section \ref{sec:summary}.
In appendixes, we explain definitions of operators and matrix elements.

\section{Definitions of TD, CD, and $E1$ operators and strengths} \label{sec:dipole}

Vortical nature of nuclear current has been discussed for a long time
(see, e.g., a review in Ref.~\cite{Kvasil:2011yk}). However, 
definition of vorticity in nuclear systems has yet to be confirmed.
To measure the nuclear vorticity,  
two different modes
have been proposed. 
One is the mode originally determined by the second order correction 
in the long-wave approximation of the transition $E\lambda$ operator in an
electromagnetic field \cite{Dubovik75,semenko81}, 
and the other is that defined based on 
multipole decomposition of the transition current density
following Ravenhall-Wambach's prescription \cite{Ravenhall:1987thb}.
In Ref.~\cite{Kvasil:2011yk}, they call the former and the latter, the toroidal and vortical modes, respectively, 
and described general treatment of toroidal, compressive, and vortical modes and their
relation to each other.
In this paper, we basically follow the descriptions  
of the TD, CD, and vortical dipole (VD) operators
in Ref.~\cite{Kvasil:2011yk}

The TD, CD, and VD operators are defined as
\begin{eqnarray}
M_\textrm{TD}(\mu)&=&\frac{-i}{2\sqrt{3}c}\int d\bvec{r} \bvec{j}(\bvec{r}) \nonumber\\
&\times& 
\left [
\frac{\sqrt{2}}{5} r^2 
\bvec{Y}_{12\mu}(\hat{\bvec{r}})+
r^2 \bvec{Y}_{10\mu} (\hat{\bvec{r}})  
\right ],\\
M_\textrm{CD}(\mu)&=&\frac{-i}{2\sqrt{3}c}\int d\bvec{r} \bvec{j}(\bvec{r}) 
 \nonumber\\
&\times& 
\left [  \frac{2\sqrt{2}}{5} r^2 \bvec{Y}_{12\mu}(\hat{\bvec{r}}) - r^2 \bvec{Y}_{10\mu} (\hat{\bvec{r}}) 
\right ],\\
M_\textrm{VD}(\mu)&=&\frac{-i}{2\sqrt{3}c}\int d\bvec{r} \bvec{j}(\bvec{r}) 
 \nonumber\\
&\times&
\left [ \frac{3\sqrt{2}}{5} r^2 
\bvec{Y}_{12\mu}(\hat{\bvec{r}})\right ],
\end{eqnarray}
where $\bvec{j}(\bvec{r})$ is the current density operator and 
$\bvec{Y}_{\lambda L\mu}$ is the vector spherical  harmonics.
Note that 
$M_\textrm{VD}=M_\textrm{TD}+M_\textrm{CD}$.
In this paper, we take into account only the convection part of the 
nuclear current but skip its magnetization (spin) part. 
The definition of $\bvec{j}(\bvec{r})$ as well as that of density $\rho(\bvec{r})$ are
given in Appendix~\ref{app:density}.
The term $\bvec{Y}_{10\mu} (\hat{\bvec{r}})$ includes the $L=1$ excitation of the 
center-of-mass motion, but it gives no contribution to the transition matrix element 
in the AMD framework because the center-of-mass motion of the 
AMD wave function is fixed to be an $S$-wave state and can be exactly removed. 

The TD operator can be written 
using a curl of the transition current density as 
$M_\textrm{TD}\propto \int d\bvec{r} 
(\nabla\times \bvec{j})\cdot (r^3 \bvec{Y}_{11\mu})$, and the CD operator, 
$M_\textrm{CD}\propto \int d\bvec{r} 
(\nabla\cdot \bvec{j})r^3Y_{1\mu}$,
is regarded as the counter part of the TD operator.
In a hydrodynamical sense, the TD and CD modes are considered to be 
vortical and irrotational, respectively. 
On the other hand, the VD operator measures the
$\bvec{Y}_{1 2 \mu}$ component of the transition current $\bvec{j}$
and free from the $\bvec{Y}_{1 0 \mu}$ component. 
In the Ravenhall-Wambash's prescription \cite{Ravenhall:1987thb}, 
$\bvec{Y}_{\lambda \lambda+1 \mu}$ and $\bvec{Y}_{\lambda \lambda-1 \mu}$ 
components of $\bvec{j}$ are interpreted as vortical and irrotational parts. In their definition, 
the VD operator is vortical, whereas  
the TD and CD operators are
mixed modes of both vortical ($\bvec{Y}_{\lambda \lambda+1 \mu}$) and irrotational 
 ($\bvec{Y}_{\lambda \lambda-1 \mu}$) components. 
Kvasil and his collaborators argued that 
the TD operator 
is a natural measure of the 
nuclear vorticity  \cite{Kvasil:2011yk,Repko:2012rj}, 
though there exist studies with the TD operator and those with the VD one. 
They demonstrated with RPA calculations 
that the TD operator is a good mode to separate 
the low-energy dipole mode from the high-energy CD mode.

For a dipole transition from the ground state, $|0\rangle \to |f\rangle$,
matrix elements of these operators 
are written with the transition current density
$\delta\bvec{j}(\bvec{r})\equiv \langle f | \bvec{j}(\bvec{r})|0 \rangle$  
as 
\begin{eqnarray}
\langle f |M_\textrm{TD}(\mu) |0 \rangle& =&
\frac{-i}{2\sqrt{3}c}\int d\bvec{r} \delta\bvec{j}(\bvec{r}) \nonumber\\
&\times& 
\left [
\frac{\sqrt{2}}{5} r^2 
\bvec{Y}_{12\mu}(\hat{\bvec{r}})+
r^2 \bvec{Y}_{10\mu} (\hat{\bvec{r}})  
\right ],\\
\langle f |M_\textrm{CD}(\mu) |0 \rangle&=&\frac{-i}{2\sqrt{3}c}\int d\bvec{r} \delta\bvec{j}(\bvec{r}) 
 \nonumber\\
&\times& 
\left [  \frac{2\sqrt{2}}{5} r^2 \bvec{Y}_{12\mu}(\hat{\bvec{r}}) - r^2 \bvec{Y}_{10\mu} (\hat{\bvec{r}}) 
\right ],\\
\langle f |M_\textrm{VD}(\mu) |0 \rangle& =&
\frac{-i}{2\sqrt{3}c}\int d\bvec{r} \delta\bvec{j}(\bvec{r})
 \nonumber\\
&\times& 
\left[\frac{3\sqrt{2}}{5} r^2 
\bvec{Y}_{12\mu}(\hat{\bvec{r}})\right].
\end{eqnarray}
By using the continuity equation 
\begin{equation}
\nabla\cdot  \bvec{j} = - \frac{i}{\hbar} \left [H,\rho \right], 
\end{equation}
the matrix element of the CD operator is straightforwardly transformed to 
that of the familiar IS dipole ($IS1$) operator as 
\begin{eqnarray}
\langle f |M_\textrm{CD}(\mu)| i \rangle &=& -\frac{1}{10} 
\frac{E}{\hbar c} \langle f |M_{IS1}(\mu)| i \rangle, \\
M_{IS1}(\mu)&\equiv&\int d\bvec{r} \rho(\bvec{r}) r^3 Y_{1\mu} (\hat{\bvec{r}}),
\end{eqnarray}
where $E$ is the excitation energy $E\equiv E_f -E_0$ given with the initial energy ($E_0$)
and final energy ($E_f$).
The $E1$ operator is written with the IV density operator
$\rho^\textrm{IV}(\bvec{r})$ as  
\begin{equation}
M_{E1}(\mu)\equiv\int d\bvec{r} \frac{1}{2}\rho^\textrm{IV}(\bvec{r})  r Y_{1\mu} (\hat{\bvec{r}}),
\end{equation}
and also written with the IV current density operator 
$\bvec{j}^\textrm{IV}({\bvec{r}})$ as 
\begin{equation}
M_{E1}(\mu) =-  \frac{i\hbar }{2E}\sqrt{\frac{3}{4\pi}} 
\int d\bvec{r}j^\textrm{IV}_\mu({\bvec{r}}).
\end{equation}

The transition strength for a dipole operator $M_\textrm{D}$ is give as
\begin{equation}
B(\textrm{D}; 0 \to f) = \frac{1}{2J_0+1}\left| \langle f |M_\textrm{D} |0 \rangle \right |^2,
\end{equation}
where $J_0$ is the angular momentum of the initial state.
We define scaled strengths of the 
TD, VD, and CD transitions
\begin{equation}
\tilde B(\textrm{TD,VD,CD}) =  \left(\frac{10\hbar c}{E}\right )^2 B(\textrm{TD,VD,CD}),
\end{equation}
so that $\tilde B(\textrm{CD})$ corresponds to the ordinary ISD strength $B(IS1)$.

\section{Dipole excitations of $^{10}$Be} \label{sec:10Be}
\subsection{Calculation scheme of sAMD+$\alpha$GCM}
We calculate the ground and $1^-$ states of $^{10}$Be with the sAMD 
combined with the $\alpha$-cluster GCM ($\alpha$GCM). 
The sAMD method with the GCM has been constructed and applied for study of
ISM, ISD, and $E1$ excitations in light nuclei such as $^{12}$C and  $^{16}$O, and neutron-rich Be
\cite{Kanada-En'yo:2013dma,Kanada-Enyo:2015knx,Kanada-Enyo:2015vwc}.
For the detailed scheme of the present calculation of $^{10}$Be, the reader is referred to the previous paper \cite{Kanada-Enyo:2015knx}. 
A similar method has been recently applied to 
study $E1$ and ISD excitations in $^{26}$Ne by Kimura \cite{Kimura:2016heo}.

In the AMD framework, a basis wave function is given by a Slater determinant,
\begin{equation}
 \Phi_{\rm AMD}({\bvec{Z}}) = \frac{1}{\sqrt{A!}} {\cal{A}} \{
  \varphi_1,\varphi_2,...,\varphi_A \},\label{eq:slater}
\end{equation}
where  ${\cal{A}}$ is the antisymmetrizer, and  $\varphi_i$ is 
the $i$th single-particle wave function written by a product of
spatial, spin, and isospin
wave functions as
\begin{eqnarray}
 \varphi_i&=& \phi_{{\bvec{X}}_i}\chi_i\tau_i,\\
 \phi_{{\bvec{X}}_i}({\bvec{r}}_j) & = &  \left(\frac{2\nu}{\pi}\right)^{3/4}
\exp\bigl[-\nu({\bvec{r}}_j-\bvec{X}_i)^2\bigr],
\label{eq:spatial}\\
 \chi_i &=& (\frac{1}{2}+\xi_i)\chi_{\uparrow}
 + (\frac{1}{2}-\xi_i)\chi_{\downarrow},
\end{eqnarray}
where $\phi_{{\bvec{X}}_i}$ and $\chi_i$ are the spatial and spin functions, respectively, and 
$\tau_i$ is the isospin
function fixed to be proton or neutron. 
The width parameter $\nu$ is chosen to be $\nu=0.19$ fm$^{-2}$ so as to minimize 
the ground state energy of $^{10}$Be.  
The condition 
\begin{equation}\label{eq:cm}
\frac{1}{A}\sum_{i=1,\ldots,A} \bvec{X}_i=0
\end{equation}
is kept for all basis AMD wave functions so that 
the center-of-mass motion can be exactly separated from the total wave function.
An AMD wave function
is specified by a set of variational parameters, ${\bvec{Z}}\equiv 
\{{\bvec{X}}_1,\ldots, {\bvec{X}}_A,\xi_1,\ldots,\xi_A \}$,
for centroids of single-nucleon Gaussian wave packets and spin orientations 
of all nucleons.

To obtain the wave function for the lowest $J^\pi$ state , we perform variation after 
projections (VAP) with the AMD wave function. 
Namely, 
the parameters ${\bvec{Z}}$
are determined by the energy variation 
after the angular-momentum and parity projections,  
\begin{eqnarray}
&& \frac{\delta}{\delta{\bvec{X}}_i}
\frac{\langle \Phi|H|\Phi\rangle}{\langle \Phi|\Phi\rangle}=0,\\
&& \frac{\delta}{\delta\xi_i}
\frac{\langle \Phi|H|\Phi\rangle}{\langle \Phi|\Phi\rangle}=0,\\
&&\Phi= P^{J\pi}_{MK}\Phi_{\rm AMD}({\bvec{Z}}),
\end{eqnarray}
where $P^{J\pi}_{MK}$ is the angular-momentum and parity projection operator. 
For $^{10}$Be, 
the variation is performed after the $J^\pi=0^+$ and  $J^\pi=1^-$ projections
to obtain the wave functions for the ground and the lowest $1^-$ states, respectively.
We denote the obtained parameter set $\bvec{Z}$ 
for the ground state as $\bvec{Z}^0_\textrm{VAP}=\{\bvec{X}^0_1,\ldots,\xi^0_{1},\ldots\}$,
and those for the $1^-_1$ state as $\bvec{Z}^{1^-_1}_\textrm{VAP}$.

To take into account 1p-1h excitations on the ground state, 
we consider small variation of single-particle wave functions of  
$\Phi_{\rm AMD}({\bvec{Z}^0_\textrm{VAP}})$ by shifting the Gaussian centroid 
of the $i$th single-particle wave function,
${\bvec{X}}^0_i\rightarrow {\bvec{X}}^0_i+\epsilon{\bvec{e}}_\sigma$, where
$\epsilon$ is an enough small constant and 
${\bvec{e}}_\sigma$ ($\sigma=1,\ldots,8$) are unit vectors for 8 directions
defined in the previous paper.
For the spin part of the shifted single-particle wave function, 
the spin-nonflip and spin-flip states given by parameters $\xi^0_i$ and 
$\bar\xi^0_i=-1/(4\xi^0_i)^*$ are adopted. In the sAMD method, 
totally $16A$ wave functions 
of the spin-nonflip and spin-flip shifted AMD wave functions with the parameters
\begin{eqnarray}
\bvec{Z}_{\rm s}^0(i,\sigma)&\equiv &
\{{\bvec{X}^0_1}',\cdots,{\bvec{X}^0_i}'+\epsilon {\bvec{e}}_\sigma,\cdots,
{\bvec{X}^0_A}', \nonumber\\
&& 
\xi^0_1,\cdots,\xi^0_i,\cdots,\xi^0_A \},\\
\bvec{Z}_{\bar{\rm s}}^0(i,\sigma)&\equiv &
\{{\bvec{X}^0_1}',\cdots,{\bvec{X}^0_i}'+\epsilon {\bvec{e}}_\sigma,\cdots,
{\bvec{X}^0_A}',\nonumber\\
&& 
\xi^0_1,\cdots,\bar\xi^0_i,\cdots,\xi^0_A \},
\end{eqnarray}
are adopted as basis wave functions in addition to the original ground state wave function 
$\Phi_{\rm AMD}({\bvec{Z}^0_\textrm{VAP}})$.
Here, we take into account the recoil effect and choose 
${{\bvec{X}}^{0}_j}'={\bvec{X}}^{0}_j-\epsilon {\bvec{e}}_\sigma/(A-1)$ 
to keep the condition \eqref{eq:cm}.

As discussed in the previous paper, 
the $^{10}$Be ground state obtained by the AMD+VAP shows a
$^6$He+$\alpha$ cluster structure with an inter-cluster distance 
$D_0=2.8$ fm even though 
any clusters are not {\it a priori} assumed in the AMD framework. 
To take into account large amplitude inter-cluster motion, we apply
the $\alpha$GCM by changing the inter-cluster distance (the $\alpha$-cluster distance from $^6$He)
$D_0\to D_0+\Delta D$. We label the parameter set 
as $\bvec{Z}^0_{\alpha}(\Delta D)$, which is specified by the shift
$\Delta D$ of the inter-cluster distance.  
The basis wave functions given by $\bvec{Z}^0_{\alpha}(\Delta D)$
($\Delta D=-1,0,1,\ldots,19,20$ fm) 
are superposed in the $\alpha$GCM.

Finally, we combine the sAMD and $\alpha$GCM by superposing 
all the basis wave functions in addition to the VAP wave functions, 
$\Phi_{\rm AMD}({\bvec{Z}}^0_\textrm{VAP})$ and $\Phi_{\rm AMD}(\bvec{Z}^{1^-_1}_\textrm{VAP})$.
Consequently, the final wave functions for the $0^+_1$ and $1^-_k$ states are given as  
\begin{eqnarray}
\Psi(J^\pi_k)&=&
\sum_K c_0(J^\pi_k; K) 
P^{J\pi}_{MK}\Phi_{\rm AMD}({\bvec{Z}}^0_\textrm{VAP})\nonumber\\
&+&\sum_K c_1(J^\pi_k;K) 
P^{J\pi}_{MK}\Phi_{\rm AMD}(\bvec{Z}^{1^-_1}_\textrm{VAP})\nonumber\\
&+&
\sum_{i=1,\ldots,A}\sum_{\sigma}\sum_K 
c_2(J^\pi_k; i,\sigma,K) \nonumber\\
&& 
\times P^{J\pi}_{MK}\Phi_{\rm AMD}({\bvec{Z}}_{\rm s}^0(i,\sigma))\nonumber\\
&+&
\sum_{i=1,\ldots,A}\sum_{\sigma}\sum_K 
c_3(J^\pi_k; i,\sigma,K) \nonumber\\
&& \times P^{J\pi}_{MK}\Phi_{\rm AMD}({\bvec{Z}}_{\bar{\rm s}}^0(i,\sigma))\nonumber\\
&+&
\sum_{\Delta D}\sum_K c_4(J^\pi_k; \Delta D,K) 
\nonumber\\
&& 
\times P^{J\pi}_{MK}\Phi_{\rm AMD}(\bvec{Z}^0_{\alpha}(\Delta D)), 
\end{eqnarray}
where coefficients $c_i$ are determined by 
diagonalization of the norm and Hamiltonian matrices.
Note that the present calculation corresponds 
to that labeled as ``sAMD+$\alpha$GCM+cfg'' in
the previous paper. 

For the dipole excitations $0^+_1\to 1^-_k$, 
the transition strength of a dipole operator $M_D$
are calculated with the obtained  sAMD+$\alpha$GCM wave functions, $\Psi(J^\pi_k)$, as
\begin{equation}
B(D; 0^+_1\to 1^-_k) =  | \langle \Psi(1^-_k)| M_D |\Psi(0^+_1)\rangle |^2. 
\end{equation}

In the present framework of the sAMD+$\alpha$GCM, 
the ground state is obtained by the VAP, and therefore, it contains 
correlations such as cluster correlations beyond mean field approximation.
Moreover, 1p-1h excitations on the ground state are taken into account in the 
sAMD model space, and also large amplitude cluster motion is treated
by means of the $\alpha$GCM. 

\subsection{Effective interactions}
The adopted effective interaction is the same as that used in the previous paper. 
It consists of the central force of the MV1 force\cite{TOHSAKI} and the
spin-orbit term of the G3RS force \cite{LS1,LS2}.
The MV1 force is given by two-range Gaussian two-body terms and a zero-range three-body term.
For parametrization of the MV1 force,  the case 1 with the Bartlett, Heisenberg, and Majorana parameters, $b=h=0$ and $m=0.62$, is used. 
As for strengths of the G3RS spin-orbit force with a two-range Gaussian form,
$u_{I}=-u_{II}\equiv u_{ls}=3000$ MeV are used. This set of interaction parameters 
describes well properties of the ground and excited states 
of $^{10}$Be and $^{12}$C with the AMD+VAP calculations
\cite{KanadaEn'yo:1998rf,KanadaEn'yo:1999ub,KanadaEn'yo:2006ze}.
For matter properties,  the MV1 force with the present parameters
gives the saturation density 
$\rho_s=0.192$ fm$^{-3}$, the saturation energy $E_s=-17.9$ MeV, 
the effective nucleon mass $m^*_\textrm{SNM}=0.59m$ for symmetric nuclear and 
$m^*_\textrm{PNM}=0.80m$ for spure neutron matters, 
the imcompressibility $K=245$ MeV, the symmetry energy $S=37.6$ MeV, and 
the slope parameter of the symmetry energy $L=47.7$ MeV. 

\subsection{Results of $^{10}$Be calculated with sAMD+$\alpha$GCM}

\subsubsection{Dipole strengths}
%%%%%%%%%%%%%%%%%%%%%%%%%%%%%%
\begin{figure}[!h]
\begin{center}
\includegraphics[width=6.5cm]{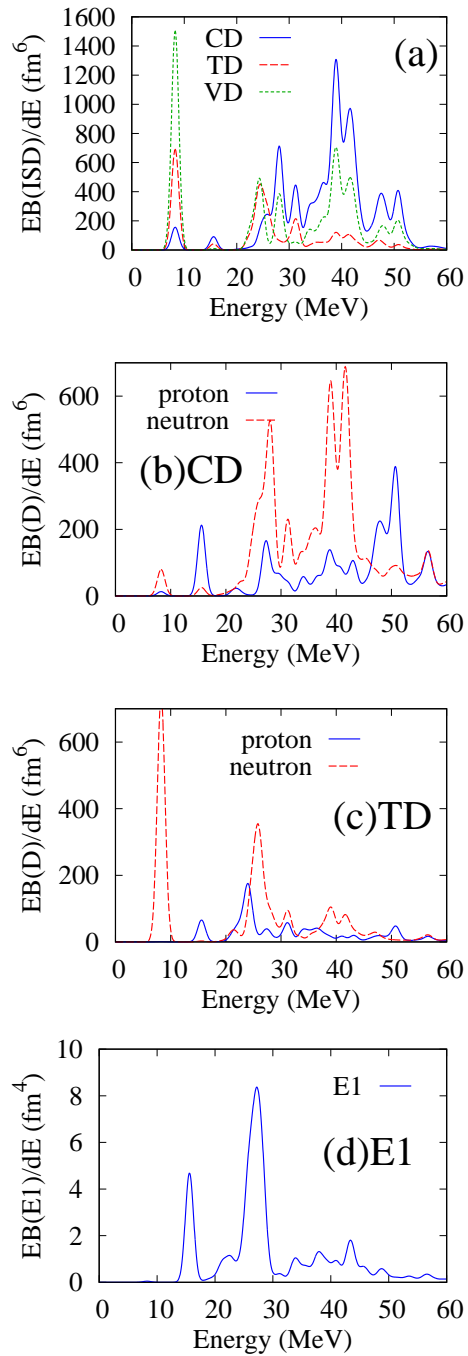} 	
\end{center}
%\vspace{0.5cm}
  \caption{(color online) 
Energy-weighted dipole strengths $E\tilde B(CD)$,  $E\tilde B(TD)$,
 $E\tilde B(VD)$, and $EB(E1)$ for the CD,TD, VD, and E1 modes of
$^{10}$Be calculated with the sAMD+$\alpha$GCM.
The proton and neutron contributions in the CD and TD modes are shown in panels 
(b) and (c). 
The smearing width is $\gamma=1$ MeV.
\label{fig:be10-ews}}
\end{figure}
%%%%%%%%%%%%%%%%%%%%%%%%%%%%%
Energy-weighted dipole strength distributions obtained with the sAMD+$\alpha$GCM 
are shown in Fig.~\ref{fig:be10-ews}.
The calculated results for ordinary IS and IV dipole, i.e., CD ($IS1$) and $E1$ strengths 
correspond to those shown in the previous paper. 
In the $E1$ excitations (see Fig.~\ref{fig:be10-ews}(d)),
a remarkable low-energy strength at $E=16$ MeV (23\% of the 
Thomas-Reiche-Kuhn sum rule)
appears below the IVGDR energy 
because of 
valence neutron motion against the $2\alpha$ core. 
The IVGDR strengths in $E\ge 20$ MeV originate in the $2\alpha$-core $E1$, namely, 
opposite oscillations between protons and neutrons in the $2\alpha$ core part.
Because of the prolate deformation of the $2\alpha$ core, the IVGDR 
shows a two-peak structure,  
the narrow peak at $E\sim 25$MeV for the the longitudinal mode and 
the broad bump around $E\sim 40$ MeV for the transverse mode, 
which is largely fragmented because of the coupling 
with the valence neutron motion.
In the CD excitations (see Fig.~\ref{fig:be10-ews}(a)), 
the broad strengths for the ISGDR are obtained 
in $E=25-50$ MeV region, relatively higher energy than the IVGDR.
Below the ISGDR, the low-energy CD strengths exhausting 5\% of 
the ISD sum rule\cite{Harakeh:1981zz}
are obtained.

ISD strengths for the TD and VD modes
are compared with the CD mode in Fig.~\ref{fig:be10-ews}(a).
In contrast to the CD mode which strongly excites
the ISGDR in the high-energy region, 
the TD strengths are dominantly distributed 
in the low-energy region rather than the high energy region.
The VD mode excites both the low-energy and high-energy dipole resonances.
It means that the TD and CD operators are suitable to separately 
probe the low-energy and high-energy parts of the ISD excitations, respectively, whereas 
the VD operator may not be a good probe to decouple the low-energy and high-energy modes.
This result is consistent with the result of the RPA calculation for $^{208}$Pb \cite{Kvasil:2011yk}.
The proton and neutron contributions in the CD and TD strengths are shown in 
Figs.~\ref{fig:be10-ews}(b) and (c). In the CD excitations, the proton contribution dominates
the strength at $E=16$ MeV and also that around $E\sim 50$ MeV, 
whereas the neutron contribution is significant for the strength around $E\sim 40$ MeV.
In the TD excitations, a remarkable strength at $E=8$ MeV comes from 
the neutron part.

Let us discuss the low-energy dipole excitations
in $E\le 20$ MeV.  Two $1^-$ resonances
are obtained at $E=8$ MeV and $E=16$ MeV. In this paper, we call 
the lower and higher ones the $1^-_1$ ($E=8$ MeV) and $1^-_2$ ($E=16$ MeV), 
which were labeled as ``B1'' and``B2'' in the previous paper, respectively.
In the $E1$ mode, 
the transition to the $1^-_1$ almost vanishes, whereas that to the $1^-_2$ is remarkably strong 
exhausting 10\% of the TRK sum rule.
In the CD mode, the strengths for both the $1^-_1$ and $1^-_2$ are not so enhanced
but visible in the strength distribution.
In the TD mode, the $1^-_1$ has a remarkably 
strong TD transition, but the $1^-_2$ shows a relatively weak TD transition. 
Thus, two low-energy dipole excitations show
 quite different transition properties;
the TD dominance in the $1^-_1$ and the $E1$ dominance in the $1^-_2$. 
%The difference between the $1^-_1$ and $1^-_2$ 
%in the dipole strength is more remarkable in the VD mode than in the TD mode; 
%the VD strength is enhanced for the $1^-_1$ and almost vanishes for the $1^-_2$. 

\section{Properties of low-energy dipole modes of $^{10}$Be} \label{sec:discussions}
As discussed previously, we obtain two low-energy dipole excitations,
the TD dominant $1^-_1$ and $E1$ dominant $1^-_2$.
Such the difference in the 
transition properties may indicate coexistence of 
two kinds of low-energy dipole modes.
In this section, we discuss properties of the low-energy dipole excitations
focusing on toroidal features. 
At first, we discuss intrinsic structures and transition current densities 
based on analysis of the AMD wave functions. Next we perform an analysis 
using a simple cluster model of $^6\textrm{He}+\alpha$ to obtain 
intuitive understanding of the dipole modes.

\subsection{Structures and transition current densities for 
the $1^-_1$, and $1^-_2$ states in the intrinsic frame}

The sAMD+$\alpha$GCM wave functions, $\Psi(0^+_1)$ and $\Psi(1^-_1)$,
for the ground and $1^-_1$ states, have more than 90\% overlap with 
the $J^\pi$-projected VAP wave functions,
$P^{0+}_{00}\Phi_{\rm AMD}({\bvec{Z}}^0_\textrm{VAP})$ and 
$P^{1-}_{MK=1}\Phi_{\rm AMD}(\bvec{Z}^{1^-_1}_\textrm{VAP})$, respectively.
Therefore 
$\Phi_{\rm AMD}({\bvec{Z}}^0_\textrm{VAP})$ and
$\Phi_{\rm AMD}({\bvec{Z}}^{1^-_1}_\textrm{VAP})$ are regarded as approximate 
intrinsic wave functions for the ground and $1^-_1$ states.
Since each AMD wave function before the projections is expressed 
by a single Slater determinant, 
we can investigate intrinsic structure of each state in the intrinsic (body-fixed) frame.
We choose the intrinsic frame $XYZ$ with the principal axes, which 
satisfy
$\langle Y^2\rangle \le \langle X^2\rangle \le \langle Z^2\rangle$
and $\langle XY\rangle = \langle YZ\rangle = \langle ZX\rangle=0$.
Here the expectation values are defined for the intrinsic state without the projections.

%%%%%%%%%%%%%%%%%%%%%%%%%%%%%%
\begin{figure}[!h]
\begin{center}
\includegraphics[width=8.0cm]{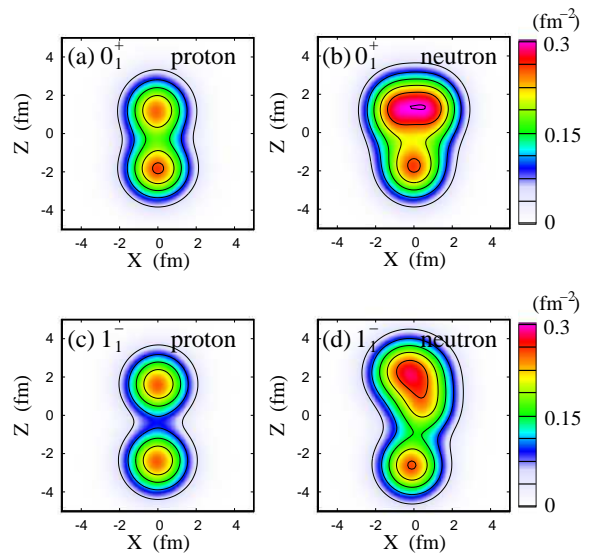} 	
\end{center}
%\vspace{0.5cm}
  \caption{(color online) 
Intrinsic density distributions of protons and neutrons in 
$^{10}\textrm{Be}(0^+_1)$ and $^{10}\textrm{Be}(1^-_1)$
obtained with the AMD+VAP calculation. Density
integrated along the $Y$ axis is plotted on the $X$-$Z$ plane.  
\label{fig:dense}}
\end{figure}
%%%%%%%%%%%%%%%%%%%%%%%%%%%%%

The intrinsic proton and neutron densities of $\Phi_{\rm AMD}({\bvec{Z}}^0_\textrm{VAP})$
and $\Phi_{\rm AMD}({\bvec{Z}}^{1^-_1}_\textrm{VAP})$ are shown in Fig.~\ref{fig:dense}.
It is found that the $0^+_1$ and $1^-_1$ show
a $2\alpha$ core with two neutrons.
One of the $\alpha$ clusters and two neutrons compose a
deformed $^6$He cluster, which is placed in the transverse 
orientation on the $Z$-axis at the $^6\textrm{He}$-$\alpha$ distance
$D=2.8$ fm in the $0^+_1$, and in the tilted (rotated) 
orientation at $D=3.9$ fm in the $1^-_1$.
It should be commented that the intrinsic state
($\Phi_{\rm AMD}({\bvec{Z}}^{1^-_1}_\textrm{VAP})$) 
with the tilted structure constructs a $K^\pi=1^-$ band consisting of 
the $J^\pi=1^-_1, 2^-_1, \ldots$ states as discussed 
in the work with the AMD+VAP \cite{KanadaEn'yo:1999ub}.

%%%%%%%%%%%%%%%%%%%%%%%%%%%%%%
\begin{figure}[!h]
\begin{center}
\includegraphics[width=6cm]{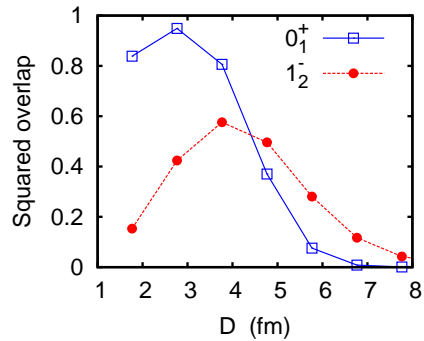} 	
\end{center}
%\vspace{0.5cm}
  \caption{(color online) 
Squared overlap of $\Psi(J^\pi_k)$ with the $\alpha$GCM basis wave functions 
$P^{J\pi}_{M0}\Phi_{\rm AMD}(\bvec{Z}^0_{\alpha}(\Delta D))$.
The overlaps for $\Psi(0^+_1)$ and $\Psi(1^-_2)$ are plotted as functions of 
the $^6\textrm{He}$-$\alpha$ distance ($D=D_0+\Delta D$). 
\label{fig:be10-over}}
\end{figure}
%%%%%%%%%%%%%%%%%%%%%%%%%%%%%

The $1^-_2$ state has significant overlap with the 
$\alpha$GCM basis wave functions
$P^{J\pi}_{MK=0}\Phi_{\rm AMD}(\bvec{Z}^0_{\alpha}(\Delta D))$
indicating that it arises from the inter-cluster ($^6\textrm{He}$-$\alpha$) excitation
 from the ground state. 
As shown in Fig.~\ref{fig:be10-over}, $\Psi(1^-_2)$ has the maximum overlap at 
$D=3.8$ fm  ($\Delta D=1$ fm) somewhat larger 
than the distance $D_0=2.8$ fm of the ground state. 
In the following analysis, we simply consider the basis AMD wave function 
$\Phi_{\rm AMD}(\bvec{Z}^0_{\alpha}(\Delta D))$ at the maximum overlap 
as an approximate intrinsic wave function for the $1^-_2$ state, though it 
has 60\% overlap with $\Psi(1^-_2)$ at most.

%%%%%%%%%%%%%%%%%%%%%%%%%%%%%%
\begin{figure}[!h]
\begin{center}
\includegraphics[width=6.0cm]{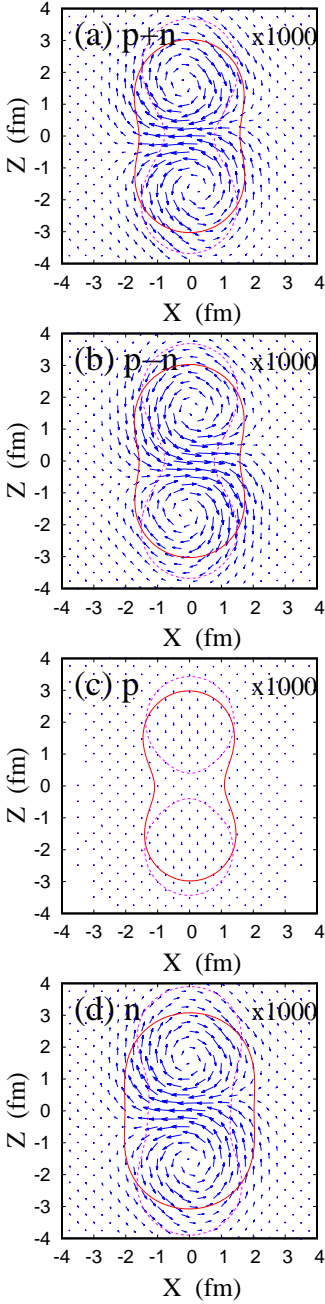} 	
\end{center}
%\vspace{0.5cm}
  \caption{(color online) 
Vector plots of the transition current densities for $|0^+_{1,\textrm{int}}\rangle \to |1^-_{1,\textrm{int}}\rangle$.
(a) IS, (b) IV, (c) proton, 
and (d) neutron transition current densities ($c$fm$^{-3}$ unit) 
at $Y=0$ are plotted on the $X$-$Z$ plane (scaled by a factor of $10^3$).
Red solid (magenta dashed) lines in the panels 
(a) and (b) show contours for the matter density $\rho(X,0,Z)=0.08$ fm$^{-3}$ 
of $|0^+_{1,\textrm{int}}\rangle$ ($|1^-_{1,\textrm{int}}\rangle$), 
and those in the panels 
(c) and (d) show contours for the proton and neutron densities 
$\rho_{p,n}(X,0,Z)=0.04$ fm$^{-3}$, respectively.
\label{fig:current-1n}}
\end{figure}
%%%%%%%%%%%%%%%%%%%%%%%%%%%%%

%%%%%%%%%%%%%%%%%%%%%%%%%%%%%%
\begin{figure}[!h]
\begin{center}
\includegraphics[width=6.0cm]{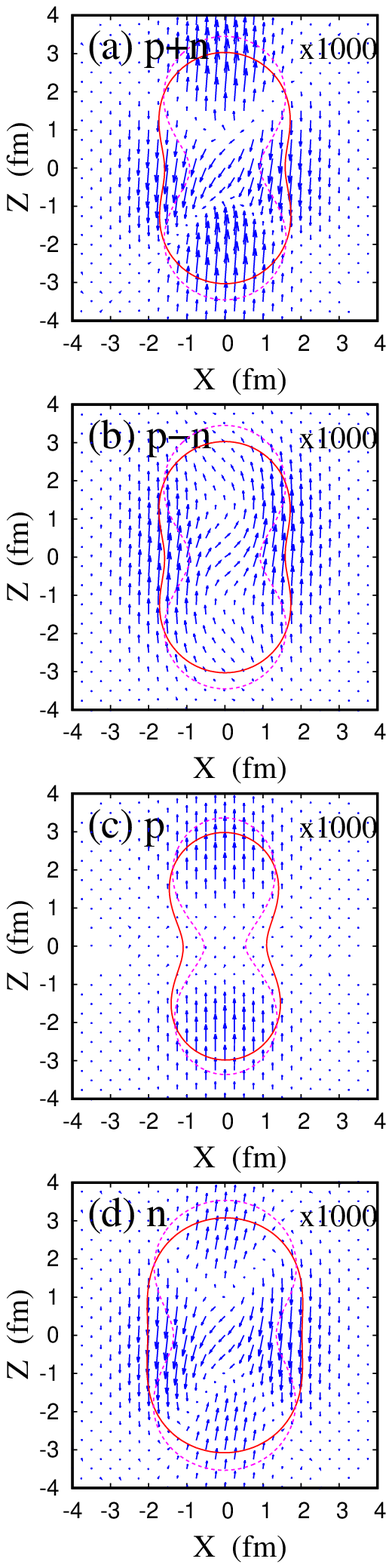} 	
\end{center}
%\vspace{0.5cm}
  \caption{(color online) 
Same as Fig.~\ref{fig:current-1n} but 
for $|0^+_{1,\textrm{int}}\rangle \to |1^-_{2,\textrm{int}}\rangle $.
\label{fig:current-1n_2}}
\end{figure}
%%%%%%%%%%%%%%%%%%%%%%%%%%%%%

We label the parity eigen states projected from 
these approximate intrinsic wave functions as 
\begin{eqnarray}
|0^+_{1,\textrm{int}}\rangle &\equiv& 
P^+\Phi_{\rm AMD}({\bvec{Z}}^0_\textrm{VAP}),\\
|1^-_{1,\textrm{int}}\rangle &\equiv& 
P^-\Phi_{\rm AMD}({\bvec{Z}}^{1^-_1}_\textrm{VAP}),\\
|1^-_{2,\textrm{int}}\rangle &\equiv& 
P^-\Phi_{\rm AMD}(\bvec{Z}^0_{\alpha}(\Delta D\textrm{=1 fm})).
\end{eqnarray}
Using the parity-projected intrinsic wave functions, 
we calculate the transition current densities,
\begin{equation}
\delta \bvec{j}(X,Y,Z)=\langle 1^-_{k, \textrm{int}}| \bvec{j}|0^+_{1,\textrm{int}}\rangle,
\end{equation}
in the intrinsic frame.
The calculated $\delta \bvec{j}(X,Y,Z)$ for $|0^+_{1,\textrm{int}}\rangle \to | 1^-_{1,\textrm{int}}\rangle$  and 
 $|0^+_{1,\textrm{int}}\rangle \to | 1^-_{2,\textrm{int}}\rangle$
at $Y=0$ on the $X$-$Z$ plane are shown in Fig.~\ref{fig:current-1n} and 
Fig.~\ref{fig:current-1n_2}.
As seen in Figs.~\ref{fig:current-1n}(a) and (d), the transition current density 
 for $|0^+_{1,\textrm{int}}\rangle \to | 1^-_{1,\textrm{int}}\rangle$  shows toroidal neutron flow 
induced by the $^6$He-cluster rotation 
(see schematic figures in Fig.~\ref{fig:he6-he4}). 
In contrast, the transition current density for $|0^+_{1,\textrm{int}}\rangle \to | 1^-_{2,\textrm{int}}\rangle$  (Fig.~\ref{fig:current-1n_2})
shows no toroidal feature but translational flow parallel to the $Z$ axis, namely, 
surface neutron flow with inner isoscalar flow caused 
by the valence neutron oscillation against the $2\alpha$ core, which is regarded as the neutron-skin oscillation mode.
(see Fig.~\ref{fig:he6-he4}(e)).

These transition current densities describe 
characteristics of the low-energy dipole excitations, i.e., the TD dominance in the $1^-_1$ and 
the $E1$ dominance in the $1^-_2$. 
In the transition $0^+_1\to 1^-_1$, 
the toroidal current gives significant contribution to the TD strength but it
gives no contribution to the $E1$ strength because it does not contain the translational mode. 
On the other hand,  in the transition $0^+_1\to 1^-_2$ arising from the 
valence neutron motion against the $2\alpha$ core,  
the $2\alpha$ motion contributes
only to the IS component but not to the IV component. Therefore, 
the surface neutron current simply enhances the $E1$ strength.
By contrast, in the IS component, 
the contribution of the surface neutron current is canceled by 
the opposite inner IS current. 
As a result of this cancellation by the recoil effect from the core, 
the TD transition is weak in $0^+_1\to 1^-_2$.

\subsection{Properties of dipole modes based on $^6$He+$\alpha$-cluster model analysis}

In the present result of $^{10}$Be, 
we obtain the remarkable $E1$ strength for the $1^-_2$ state 
because of the valence neuron motion against the $2\alpha$ core. This
corresponds to the neutron-skin oscillation mode, which has been expected to appear
in low-energy $E1$ strength of neutron-rich nuclei.
For the $1^-_2$ state, the TD strength almost vanishes because of the cancellation 
of the surface neutron current and the inner IS current of the recoiled core. 
The vanishing of the TD strength is not trivial because 
the neutron-skin oscillation mode could contain some toroidal component through 
the neutron flow along the surface. 
Indeed, the transition current density in $0^+_1\to 1^-_2$ shows such the surface 
neutron flow, which is naively expected to somewhat contribute to the TD strength if the opposite 
contribution from the recoiled core is absent. 
Moreover, there is no obvious reason why the CD strength for the $1^-_2$ state is visible 
in the CD strength distribution compared with the TD strength.
Unfortunately, at a glance on the transition current densities,  it is not easy to understand quantitatively the cancellation between the valence neutron and core contributions in the TD strength.

%%%%%%%%%%%%%%%%%%%%%%%%%%%%%%
\begin{figure}[!h]
\begin{center}
\includegraphics[width=8.0cm]{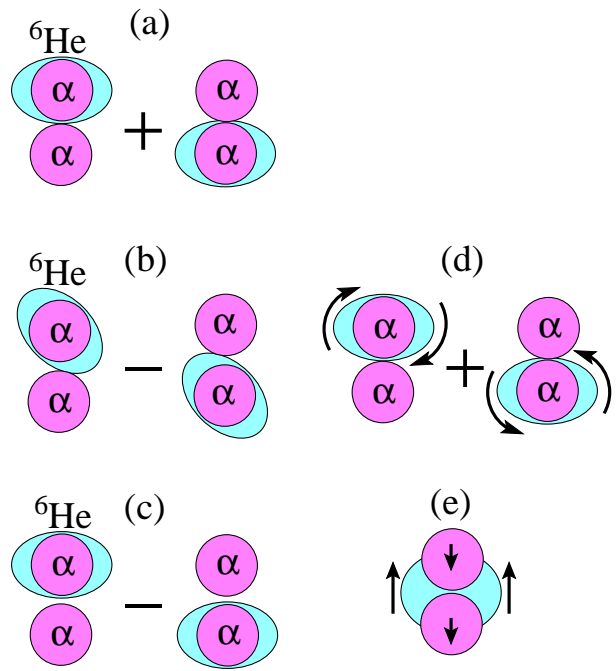} 	
\end{center}
%\vspace{0.5cm}
  \caption{(color online) 
Schematic figures of (a) $^{10}\textrm{Be}(0^+_1)$, 
(b) $^{10}\textrm{Be}(1^-_1)$, and (c) $^{10}\textrm{Be}(1^-_2)$, 
(d) transition current in $0^+_1\to 1^-_1$, and (e) that in  $0^+_1\to 1^-_2$.
\label{fig:he6-he4}}
\end{figure}
%%%%%%%%%%%%%%%%%%%%%%%%%%%%%

%%%%%%%%%%%%%%%%%%%%%%%%%%%%%%
\begin{figure}[!h]
\begin{center}
\includegraphics[width=6.0cm]{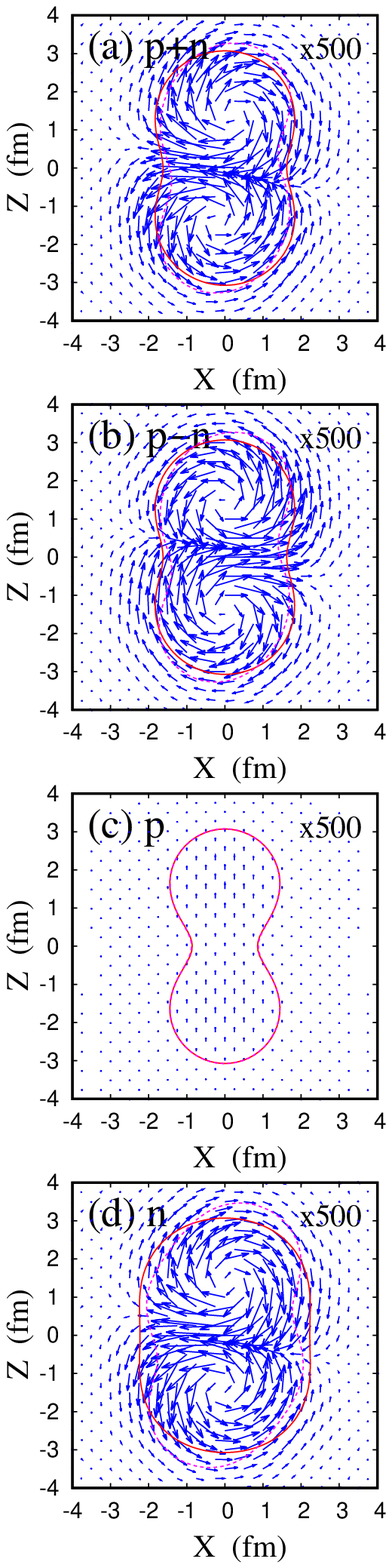} 	
\end{center}
%\vspace{0.5cm}
  \caption{(color online) 
Same as Fig.~\ref{fig:current-1n} but calculated for $C^+_\textrm{T}\to C^-_\textrm{R}$
with the $^6\textrm{He}+\alpha$ cluster model at $D=3$ fm.
The current densities ($c$fm$^{-3}$ unit) are scaled by a factor of $500$. 
\label{fig:current-45}}
\end{figure}
%%%%%%%%%%%%%%%%%%%%%%%%%%%%%

%%%%%%%%%%%%%%%%%%%%%%%%%%%%%%
\begin{figure}[!h]
\begin{center}
\includegraphics[width=6.0cm]{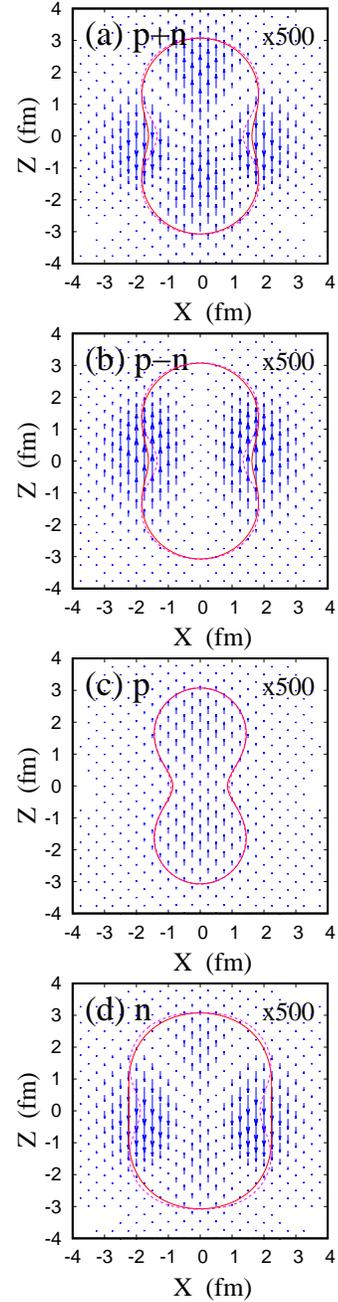} 	
\end{center}
%\vspace{0.5cm}
  \caption{(color online) 
Same as Fig.~\ref{fig:current-1n}  but calculated for $C^+_\textrm{T}\to C^-_\textrm{T}$
with the $^6\textrm{He}+\alpha$ cluster model at $D=3$ fm.
The current densities ($c$fm$^{-3}$ unit) are scaled by a factor of $500$. 
\label{fig:current-90}}
\end{figure}
%%%%%%%%%%%%%%%%%%%%%%%%%%%%%

To discuss essential properties of the TD and CD components in $0^+_1\to 1^-_1$ and 
$0^+_1\to 1^-_2$, we simply consider 
$^6$He+$\alpha$-cluster model wave functions instead of the AMD wave functions 
and analyze detailed contributions of transition current density to the TD and CD strengths in the
strong coupling picture.
We here introduce the parity-projected 
Brink-Bloch (BB) cluster wave functions of
$^6\textrm{He}+\alpha$ clustering, 
\begin{eqnarray}
\Phi^{\pi}_{\textrm{BB}} (\beta;D) &=& \frac{1}{n_0}P^\pi{\cal A}\left[ 
\Phi^\beta_{^6\textrm{He}}(\bvec{S}_1)\Phi_\alpha 
(\bvec{S}_2)  \right],\\
\bvec{S}_1&=&(0,0,\frac{2}{5}D), \\
\bvec{S}_2&=&(0,0,-\frac{3}{5}D),
\end{eqnarray}
where $n_0$ is the normalization factor determined 
by the condition $|\Phi^{\pi}_{\textrm{BB}} (\beta;D)|=1$, and 
$\Phi^\beta_{^6\textrm{He}}(\bvec{S})$ and $\Phi_{\alpha}(\bvec{S})$
are $^6\textrm{He}$- and $\alpha$-cluster wave functions given by 
the harmonic oscillator (h.o.) shell model 
$(0s)^4 p^2$ and $(0s)^4$ configurations
localized around the position $\bvec{S}$. $\beta$ is the label for the valence neutron
configuration in $p$ shell. We label $p^2_x$ (transverse) 
configuration as $\beta=C_\textrm{T}$, and $[(p_x-p_z)/\sqrt{2}]^2$ (rotated)
one as $\beta=C_\textrm{R}$. 
In short, we denote $\Phi^{+}_{\textrm{BB}} (C_\textrm{T})$,
$\Phi^{-}_{\textrm{BB}} (C_\textrm{R})$, and   
$\Phi^{-}_{\textrm{BB}} (C_\textrm{T})$ as ``$C^+_\textrm{T}$'', 
``$C^-_\textrm{R}$'', and ``$C^-_\textrm{T}$'', 
which correspond   
to the intrinsic states of the $0^+_1$,   $1^-_1$,  and  $1^-_2$,  respectively.
Schematic figures for these three configurations are drawn in Figs.~\ref{fig:he6-he4}(a), (b), and (c).
In the present analysis, we take the h.o. width $\nu=0.19$ fm$^{-1}$ and the inter-cluster distance
$D=3$ fm.

These wave functions have planer configurations restricted on the $X$-$Z$ plane,
and therefore they are suitable to discuss essential features of the transition current densities
projected onto the $X$-$Z$ plane.
The transition current densities for $C^+_\textrm{T}\to C^-_\textrm{R}$ and 
$C^+_\textrm{T}\to C^-_\textrm{T}$ are shown in Fig.~\ref{fig:current-45}
and Fig.~\ref{fig:current-90}, respectively.
The transition, $C^+_\textrm{T}\to C^-_\textrm{R}$, shows
the toroidal current similar to that found in 
$|0^+_{1,\textrm{int}}\rangle \to | 1^-_{1,\textrm{int}}\rangle$, and  
$C^+_\textrm{T}\to C^-_\textrm{T}$ shows
the translational current similarly to 
$|0^+_{1,\textrm{int}}\rangle \to | 1^-_{2,\textrm{int}}\rangle$.

Let us discuss detailed contributions of the transition current densities
to the TD and CD modes in the intrinsic frame. For this aim, we define 
$K$ components of the TD and CD operators in the frame, 
$M_\textrm{TD,CD}(K=0)$ and $M_\textrm{TD,CD}(|K|=1)$. 
The definition and explicit notation of $M_\textrm{TD,CD}(K=0)$ 
and $M_\textrm{TD,CD}(|K|=1)$ are given in Appendix \ref{app:transitions}. 
In the strong coupling picture, 
the TD and CD transition matrix elements are proportional to 
the integration of the following TD and CD transition
densities at $Y=0$ on the $X$-$Z$ plane,
\begin{eqnarray}
{\cal M}^{K=0}_\textrm{TD}(X,0,Z)&&= \frac{-i}{20c} \sqrt{\frac{3}{\pi}}
\nonumber\\
&& \times
 \left(2 X^2\delta j_Z + Z^2 \delta j_Z - ZX \delta j_X\right),\label{eq:TD0}\\
{\cal M}^{K=0}_\textrm{CD}(X,0,Z)&&=\frac{-i}{20c} \sqrt{\frac{3}{\pi}}
\nonumber\\
&& \times
 \left(-X^2\delta j_Z -3 Z^2 \delta j_Z -2 ZX \delta j_X\right),\label{eq:CD0}\nonumber\\
\\
{\cal M}^{|K|=1}_\textrm{TD}(X,0,Z)&&= \frac{-i}{20c} \sqrt{\frac{3}{\pi}}
\nonumber\\
&& \times
\left( 2 Z^2\delta j_X + X^2\delta j_X -XZ\delta j_Z \right),\label{eq:TD1}\nonumber\\\\ 
{\cal M}^{|K|=1}_\textrm{CD}(X,0,Z)&&=\frac{-i}{20c} \sqrt{\frac{3}{\pi}}
\nonumber\\
&& \times
\left(- Z^2\delta j_X - 3 X^2\delta j_X -2XZ\delta j_Z \right)\label{eq:CD1},\nonumber\\
\end{eqnarray}
meaning that ${\cal M}^{K}_\textrm{TD,CD}$ can be 
decomposed into three terms.

\begin{figure}[!h]
\begin{center}
\includegraphics[width=8.0cm]{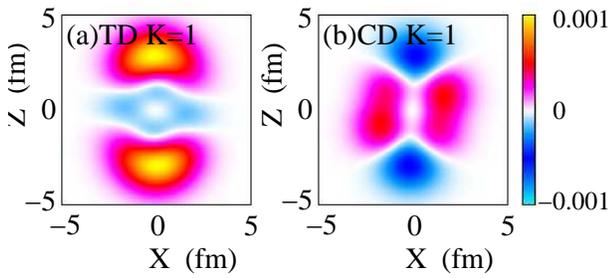} 	
\end{center}
%\vspace{0.5cm}
  \caption{(color online) 
$|K|=1$ component of (a) TD and (b) CD transition densities 
 (($-i$)cfm$^{-1}$) at $Y=0$ on the 
$X$-$Z$ plane for $C^+_\textrm{T}\to C^-_\textrm{R}$.
\label{fig:tmcm1-45}}
\end{figure}

\begin{figure}[!h]
\begin{center}
\includegraphics[width=8.0cm]{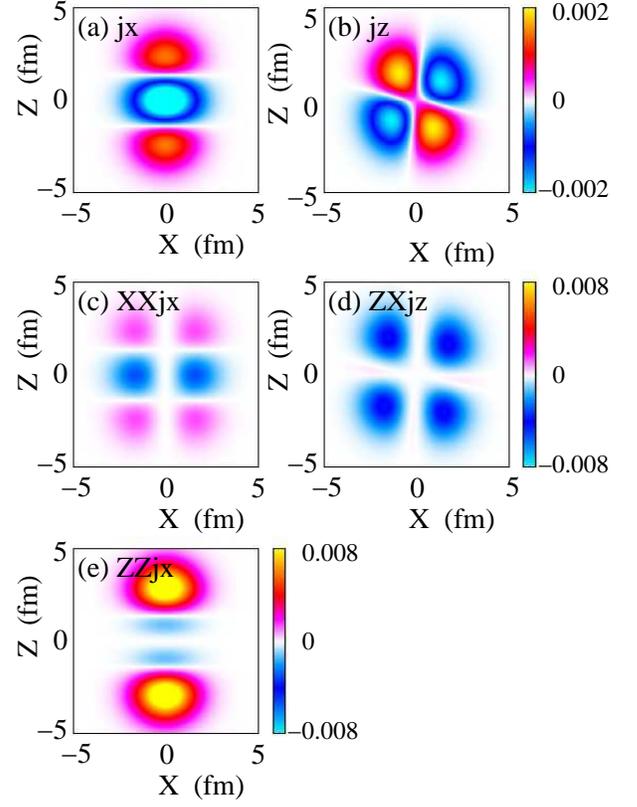} 	
\end{center}
%\vspace{0.5cm}
  \caption{(color online) 
Decomposition of the
$|K|=1$ component of the 
transition current densities
for $C^+_\textrm{T}\to C^-_\textrm{R}$.
Non-weighted transition current densities, (a) $j_X$ and (b) $j_Z$ (cfm$^{-3}$), and 
the weighted transition current densities, 
(c) $X^2j_X$, (d) $XZj_Z$, and (e) $Z^2j_X$  (cfm$^{-1}$),  at $Y=0$ are plotted on the 
$X$-$Z$ plane.
\label{fig:tmcm1-45-part}}
\end{figure}

For $C^+_\textrm{T}\to C^-_\textrm{R}$ corresponding to
$0^+_1\to 1^-_1$, we consider the $K=1$ transition
because of the $K=1$ feature of the $1^-_1$ state. As given in 
\eqref{eq:TD1}, the ratio of weight factors of 
three terms $Z^2\delta j_X$, $X^2\delta j_X$, and $XZ\delta j_Z$
in  the TD mode is $2:1:-1$, which indicates that 
$Z^2\delta j_X$ is the major term. 
Similarly, $X^2\delta j_X$ is the major term in the $K=1$ CD mode.
The $K=1$ component of the TD and CD transition densities 
for $C^+_\textrm{T}\to C^-_\textrm{R}$
and its decomposition are shown in Fig.~\ref{fig:tmcm1-45} and
Fig.~\ref{fig:tmcm1-45-part}. 
Clearly seen in the decomposition, the TD transition is enhanced 
because of remarkable contribution from the $Z^2j_X$ term (Fig.~\ref{fig:tmcm1-45-part}(e)). In addition, 
the $XZ\delta j_Z$ term (Fig.~\ref{fig:tmcm1-45-part}(d)) gives coherent contribution to the TD transition. Consequently, the TD transition is remarkably strong for $0^+_1\to 1^-_1$.
However, the CD transition is not enhanced because of the  
cancellation between positive and negative contributions in the major term, $X^2\delta j_X$
(Fig.~\ref{fig:tmcm1-45-part}(c)).
Moreover, further cancellation is caused by decoherent contribution from the $Z^2j_X$ term
(Fig.~\ref{fig:tmcm1-45-part}(e)).

%%%%%%%%%%%%%%%%%%%%%%%%%%%%%%
%%%%%%%%%%%%%%%%%%%%%%%%%%%%%

%%%%%%%%%%%%%%%%%%%%%%%%%%%%%%
%%%%%%%%%%%%%%%%%%%%%%%%%%%%%

\begin{figure}[h]
\begin{center}
\includegraphics[width=8.0cm]{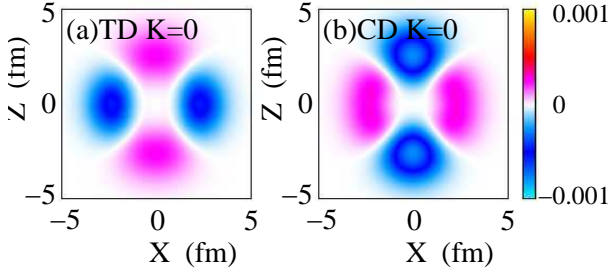} 	
\end{center}
%\vspace{0.5cm}
  \caption{(color online) 
$K=0$ component of (a) TD and (b) CD transition densities 
 (($-i$)cfm$^{-1}$) at $Y=0$ on the 
$X$-$Z$ plane for $C^+_\textrm{T}\to C^-_\textrm{T}$.
%(b)(e) Proton and (c)(f) neutron components are also shown.
\label{fig:tmcm0-90}}
\end{figure}

\begin{figure}[h]
\begin{center}
\includegraphics[width=8.0cm]{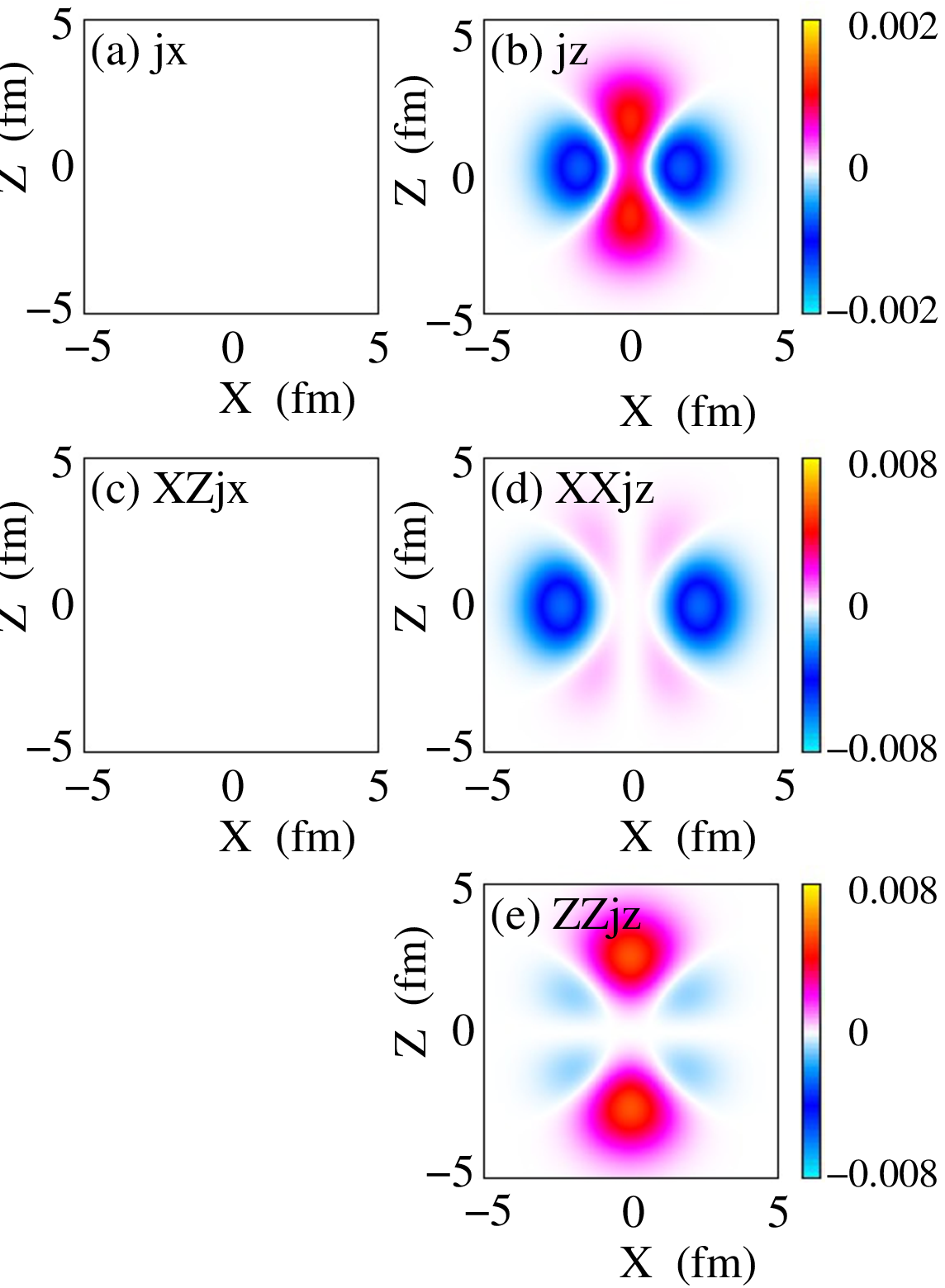} 	
\end{center}
%\vspace{0.5cm}
  \caption{(color online) 
Decomposition of the
$K=0$ component of the 
transition current densities
for the transition $C^+_\textrm{T}\to C^-_\textrm{T}$.
Non-weighted transition current densities, (a) $j_X$ and (b) $j_Z$ (cfm$^{-3}$), and 
the weighted transition current densities, 
(c) $ZXj_X$, (d) $XX^2j_Z$, and (e) $ZZj_Z$  (cfm$^{-1}$),  at $Y=0$ are plotted on the 
$X$-$Z$ plane. 
\label{fig:tmcm0-90-part}}
\end{figure}

For $C^+_\textrm{T}\to C^-_\textrm{T}$ corresponding to
$0^+_1\to 1^-_2$, we consider the $K=0$ transition
because of the $K=0$ feature of the $1^-_2$ state.
As given in 
\eqref{eq:TD0} and \eqref{eq:CD0}, the major terms of the TD and CD  modes are 
 $X^2\delta j_Z$ and $Z^2\delta j_Z$, respectively. 
The $K=0$ component of the TD and CD transition densities 
for 
$C^+_\textrm{T}\to C^-_\textrm{T}$ and its decomposition 
are shown in Fig.~\ref{fig:tmcm0-90} and
Fig.~\ref{fig:tmcm0-90-part}. 
As seen in Fig.~\ref{fig:tmcm0-90}(a), 
the TD transition almost vanishes because of the cancellation between 
positive contribution from the $2\alpha$ core part and the negative contribution from 
valence neutrons in the TD transition density. 
In the decomposition of the TD transition density, 
the $X^2J_Z$ and $Z^2J_Z$ terms
cancel each other, whereas the $ZXJ_X$ term vanishes itself. Note that 
the factor $X^2$ in $X^2J_Z$ enhances the surface neutron contribution
and the factor $Z^2$ in $Z^2J_Z$ enhances the $2\alpha$ core contribution.
It means that the TD transition is suppressed because the contribution of the 
surface neutron current is canceled by the recoil effect of the $2\alpha$ core.
Also in the CD transition, the $Z^2J_Z$ (surface neutron) contribution is somewhat 
canceled by the $X^2J_Z$ (core) contribution. 
However, since the $Z^2J_Z$ term dominates the $K=0$ component of the CD mode
with the factor of 3, the surface neutron contribution remains in the CD mode. As a result, 
the neutron-skin oscillation mode in $0^+_1\to 1^-_2$ contains 
almost no TD component but some CD component.

%%%%%%%%%%%%%%%%%%%%%%%%%%%%%%
%%%%%%%%%%%%%%%%%%%%%%%%%%%%%

%%%%%%%%%%%%%%%%%%%%%%%%%%%%%%
%%%%%%%%%%%%%%%%%%%%%%%%%%%%%

%%%%%%%%%%%%%%%%%%%%%%%%%%%%%%
\begin{figure}[!h]
\begin{center}
\includegraphics[width=7.0cm]{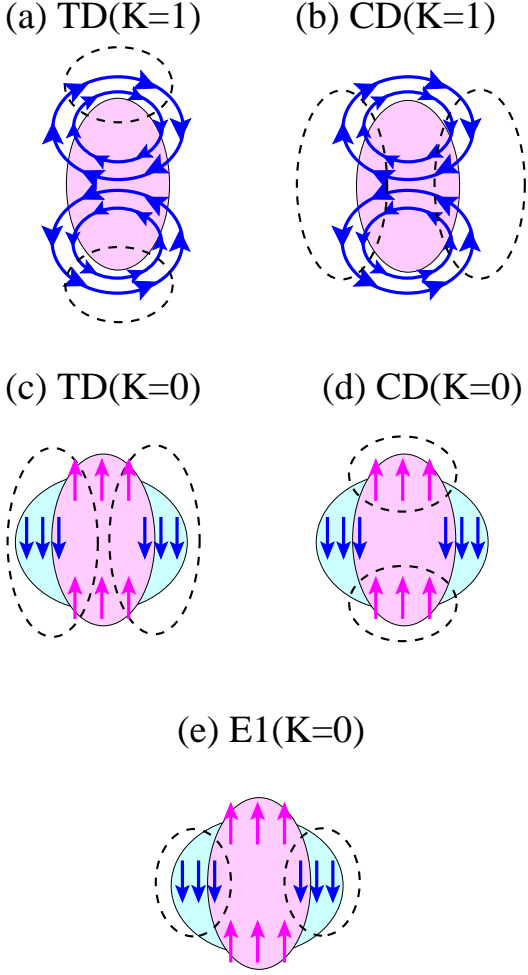} 	
\end{center}
%\vspace{0.5cm}
  \caption{(color online) 
Schematic figures for transition current densities and their contributions
to TD, CD, and $E1$ strengths. 
(a)(b) Contributions of the toroidal current to $K=1$ component of the TD and CD modes
in $0^+_1\to 1^-_1$. 
(c)(d)(e) Contributions of the translational current to $K=0$ component of the TD, CD and $E1$
modes in $0^+_1\to 1^-_2$. 
\label{fig:current-fig}}
\end{figure}
%%%%%%%%%%%%%%%%%%%%%%%%%%%%%

Based on the present analysis of the transition current densities in the intrinsic frame, 
we can obtain the following understanding of 
transition current contributions to the TD, CD, and $E1$ strengths. 
The $0^+_1\to 1^-_1$ transition is characterized by the toroidal neutron current.
In particular, the surface neutron current in the longitudinal part 
gives significant contribution to the 
$K=1$ TD  mode (Fig.~\ref{fig:current-fig}(a)). On the other hand, the toroidal neutron current
gives small contribution to the $K=1$ CD mode
because of the cancellation in the side region (Fig.~\ref{fig:current-fig}(b)).
The $0^+_1\to 1^-_2$ transition is characterized by the 
surface neutron and inner IS currents along the $Z$ axis, which are induced by opposite oscillation between
valence neutrons and the $2\alpha$ core. 
In the $K=0$ TD mode, the contribution of the inner IS current from the core part cancels that 
of the surface neutron current in the side region (Fig.~\ref{fig:current-fig}(c)).
In the $K=0$ CD mode, the longitudinal part of the IS current from the recoiled core 
gives some contribution (Fig.~\ref{fig:current-fig}(d)).
In the $E1$ mode, the surface neutron current along the $Z$ axis simply contributes to the $E1$
strength without cancellation by the core IS current (Fig.~\ref{fig:current-fig}(e)).

It should be commented that the BB cluster model wave functions 
reach to shell-model wave functions in a small limit of the inter-cluster distance.
The shell model limit of $C^+_\textrm{T}$ is the $(000)^4(001)^4 (100)^2$ configuration.
Here $(n_x n_y n_z)$ is the notation for oscillator quanta of single-particle orbits in a  
h.o. potential.
The shell model limit of $C^-_\textrm{R}$ is dominated by the 
$(000)^4(001)^4 (100) (002)$ configuration, 
which corresponds to the 1p-1h excitation, $(100)^{-1} (002)^1$ on 
the vacuum  state $(000)^4(001)^4 (100)^2$. 
Strictly speaking, it also contains additional 1p-1h configurations 
because of the recoil effect along the $Z$ axis, but they do not contribute to the $|K|=1$
dipole excitations. 
The shell model limit of $C^-_\textrm{T}$ is expressed by a linear combination 
of $(100)^{-1} (101)^1$ and $(001)^{-1} (002)^1$ excitations on the vacuum state.
The former is neutron excitation, and 
the latter is the core recoil effect
and gives IS contributions (coherent proton and neutron contributions).

\section{Summary and outlook} \label{sec:summary}

We studied dipole excitations in $^{10}$Be based on the sAMD+$\alpha$GCM calculation.
The present model takes into account 1p-1h excitations and 
large amplitude cluster modes, and is useful  to describe low-energy dipole strengths for 
cluster modes and high-energy ones for GDRs.
It should be stressed that 
the parity and angular-momentum projections are performed and the center-of-mass motion 
can be exactly separated in the present model.

By calculations of the TD, CD, $E1$ strengths,  
the toroidal and compressive properties of the ISD excitations as well as the $E1$ 
property have been investigated. It was found that the TD and CD modes
dominate the low-energy and high-energy parts of the ISD strengths, respectively. 
It indicates that the toroidal operator is a good mode to probe  
the low-energy dipole resonances separately from the ISGDR.

In the low-energy region $E\le 20$ MeV, we obtained two $1^-$ states, 
the TD dominant $1^-_1$ state at $E=8$ MeV and the $E1$ dominant 
$1^-_2$ state at $E=16$ MeV. 
The $0^+_1$,  $1^-_1$, and $1^-_2$ states have cluster structures 
with a $2\alpha$ core and two neutrons regarded as the $^6\textrm{He}+\alpha$ clustering. 
The $0^+_1\to 1^-_1$ excitation is interpreted as 
rotation mode of the deformed $^6$He cluster, whereas the $0^+_1\to 1^-_2$ excitation 
arises from the inter-cluster ($^6\textrm{He}$-$\alpha$) motion.  
The transition current density for $0^+_1\to 1^-_1$ shows the toroidal neutron flow 
caused by the $^6$He-cluster rotation.
In contrast, that for $0^+_1\to 1^-_2$
shows no toroidal feature but the surface neutron flow with the inner IS flow caused 
by the surface neutron oscillation against the $2\alpha$ core, i.e., 
the neutron-skin oscillation mode. These properties of transition current densities describe 
the TD dominance in the $1^-_1$ and the $E1$ dominance in the $1^-_2$. 

In $^{10}$Be, valence neutron motion around the axial symmetric core
is essential for the low-energy dipole modes. 
The coexistence of two different modes, the toroidal mode and the neutron-skin oscillation one, 
in the  low-energy region might be related to decoupling of the $K=1$ and $K=0$ 
dipole modes in the deformed system.
An interesting question is whether such phenomena generally appear in deformed neutron-rich nuclei.
It is also interesting to search for TD dominant states 
caused by rotation of a deformed cluster in other nuclei. 
Experimentally, there is no established method to directly measure TD strengths in unstable nuclei.
Analysis of hadronic scatterings based on a reliable reaction theory might be a promising tool.

%\section*{Acknowledgments} 
\begin{acknowledgments}
The computational calculations of this work were performed by using the
supercomputer in the Yukawa Institute for theoretical physics, Kyoto University. This work was supported by 
JSPS KAKENHI Grant Number 26400270.
\end{acknowledgments}

\appendix

\section{density and current density operators}\label{app:density}
The density and current density operators for nuclear matter are defined as  
\begin{eqnarray}
\rho(\bvec{r})&=& \sum_k  \delta(\bvec{r}-\bvec{r}_k),\\
\bvec{j}(\bvec{r})&=& -\frac{i\hbar}{2m} \sum_k  \nabla_k\delta(\bvec{r}-\bvec{r}_k)+\delta(\bvec{r}-\bvec{r}_k)\nabla_k.
\end{eqnarray}  
The current considered here is the convective part of the nuclear current but does not contain
the magnetization (spin) part.
The transition current density for $|0\rangle \to |f\rangle$ is defined as
\begin{equation}
\delta\bvec{j}(\bvec{r})= \langle f |\bvec{j}(\bvec{r}) |0 \rangle.
\end{equation}  
The IV, proton, and neutron components are
\begin{eqnarray}
\rho^\textrm{IV}(\bvec{r})&= & \rho^p(\bvec{r})-\rho^n(\bvec{r}),\\
\rho^{p(n)}(\bvec{r})&=& \sum_{k\in p(n)}  \delta(\bvec{r}-\bvec{r}_k),\\
\bvec{j}^\textrm{IV}(\bvec{r})&=& 
 \bvec{j}^p(\bvec{r})-\bvec{j}^n(\bvec{r}),\\
\bvec{j}^{p(n)}&=&
-\frac{i\hbar}{2m} \sum_{k \in p(n) }
\nonumber\\
&&\times
\left[\nabla_k\delta(\bvec{r}-\bvec{r}_k)+\delta(\bvec{r}-\bvec{r}_k)\nabla_k\right],\\
\delta\bvec{j}^\textrm{IV}(\bvec{r})&=& 
\delta\bvec{j}^p(\bvec{r})- \delta\bvec{j}^n(\bvec{r}),\\
\delta\bvec{j}^{p(n)}(\bvec{r})&= &\langle f |\bvec{j}^{p(n)}(\bvec{r}) |0 \rangle.
\end{eqnarray}

\section{Transitions in intrinsic frame}\label{app:transitions}
We define the $K=0$ and $K=1$ components of the 
TD and CD operators in the intrinsic (body-fixed) frame, $XYZ$,  as
\begin{eqnarray}
M_\textrm{TD,CD}(K=0)&\equiv& M_\textrm{TD,CD}(\mu=0),\\
M_\textrm{TD,CD}(|K|=1)&\equiv& -\frac{1}{\sqrt{2}}
\left[M_\textrm{TD,CD}(\mu=+1)
\right.\nonumber\\ && \left.
-M_\textrm{TD,CD}(\mu=+1)\right].
\end{eqnarray}
In the strong-coupling picture, 
the $K=0$ and $|K|=1$ transitions for $|0_\textrm{int}\rangle\to |f_\textrm{int}\rangle$
are expressed as  
\begin{eqnarray}
&&\langle f_\textrm{int} | M_\textrm{TD,CD}(K=0, |K|=1) | 0_\textrm{int} \rangle 
\nonumber\\
&&\qquad \qquad\qquad =\int d\bvec{r}_\textrm{int} {\cal M}^{K=0, |K|=1}_\textrm{TD,CD}(\bvec{r}_\textrm{int}),
\end{eqnarray}
with
\begin{eqnarray}
{\cal M}^{K=0}_\textrm{TD}(\bvec{r}_\textrm{int})&=& \frac{-i}{20c} \sqrt{\frac{3}{\pi}}\left[ 2(X^2+Y^2)\delta j_Z + Z^2\delta j_Z 
\right.\nonumber\\ && \left.
-ZX\delta j_X-ZY\delta j_Y \right] ,\\
{\cal M}^{K=0}_\textrm{CD}(\bvec{r}_\textrm{int})&=&\frac{-i}{20c} \sqrt{\frac{3}{\pi}}
 \left[ -(X^2+Y^2)\delta j_Z -3 Z^2 \delta j_Z 
\right.\nonumber\\ && \left.
-2 ZX\delta j_X-2 ZY\delta j_Y \right] ,\\
{\cal M}^{|K|=1}_\textrm{TD}(\bvec{r}_\textrm{int})&=& \frac{-i}{20c} \sqrt{\frac{3}{\pi}}\left[ 2(Y^2+Z^2)\delta j_X + X^2\delta j_X
\right.\nonumber\\ && \left.
 -XY\delta j_Y-XZ\delta j_Z \right] ,\\
{\cal M}^{|K|=1}_\textrm{CD}(\bvec{r}_\textrm{int})&=&\frac{-i}{20c} \sqrt{\frac{3}{\pi}}
 \left[ -(Y^2+Z^2)\delta j_X -3 X^2 \delta j_X 
\right.\nonumber\\ && \left.
-2 XY\delta j_Y-2 XZ\delta j_Z \right] ,
\end{eqnarray}
where $\bvec{r}_\textrm{int}=(X,Y,Z)$. 
We call ${\cal M}_\textrm{TD(CD)}$ the TD(CD) transition 
density. 
For $j_Y=0$ case, the TD and CD transition densities at $Y=0$ are written as 
\begin{eqnarray}
{\cal M}^{K=0}_\textrm{TD}(X,0,Z)&=& \frac{-i}{20c} \sqrt{\frac{3}{\pi}}
\nonumber\\&\times&
 \left(2 X^2\delta j_Z + Z^2 \delta j_Z - ZX \delta j_X\right) ,\nonumber\\
\\
{\cal M}^{K=0}_\textrm{CD}(X,0,Z)&=&\frac{-i}{20c} \sqrt{\frac{3}{\pi}}
\nonumber\\&\times&
 \left(-X^2\delta j_Z -3 Z^2 \delta j_Z -2 ZX \delta j_X\right) ,\nonumber\\
\\
{\cal M}^{|K|=1}_\textrm{TD}(X,0,Z)&=& \frac{-i}{20c} \sqrt{\frac{3}{\pi}}
\nonumber\\&\times&
\left( 2 Z^2\delta j_X + X^2\delta j_X -XZ\delta j_Z \right),\nonumber\\
\\ 
{\cal M}^{|K|=1}_\textrm{CD}(X,0,Z)&=&\frac{-i}{20c} \sqrt{\frac{3}{\pi}}
\nonumber\\&\times&
\left(- Z^2\delta j_X - 3 X^2\delta j_X -2XZ\delta j_Z \right).\nonumber\\
\end{eqnarray}

\section{Vector spherical harmonics}
For given vectors $\bvec{a}$ and $\bvec{b}$, 
\begin{eqnarray}
\hat{\bvec{a}}\cdot \bvec{Y}_{\lambda L\mu}(\hat{\bvec{b}})&=&\sqrt{\frac{4\pi}{3}} \left[ Y_ L(\hat{\bvec{b}})\otimes Y_1(\hat{\bvec{a}})  \right]_{\lambda\mu},\\ 
 \left[ Y_ L(\hat{\bvec{b}})\otimes Y_l(\hat{\bvec{a}})  \right]_{\lambda\mu}&\equiv&
\sum_{M,m}\langle LMlm|\lambda\mu \rangle 
\nonumber\\
&\times&
Y_ {LM}(\hat{\bvec{b}})Y_{lm}(\hat{\bvec{a}}).
\end{eqnarray}
Explicit expressions of $\bvec{a}\cdot[r^{\lambda+1}\bvec{Y}_{\lambda L\mu}(\hat{\bvec{r}})]$ 
for $\lambda=1$ and $\mu=0$ are 
\begin{eqnarray}
\bvec{a}\cdot \left[r^2\bvec{Y}_{100}(\hat{\bvec{r}})\right]&=&
\sqrt{\frac{4\pi}{3}} a r^2 Y_{00}(\hat{\bvec{r}}) Y_{10}(\hat{\bvec{a}})\nonumber\\ 
&=&\frac{1}{\sqrt{4\pi}}
\left(x^2 +y^2+z^2\right) a_z ,\\
\bvec{a}\cdot \left[r^2\bvec{Y}_{120}(\hat{\bvec{r}})\right]&=&
\sqrt{\frac{4\pi}{3}} a r^2 
\left[ 
\sqrt{\frac{3}{10}}Y_{21}(\hat{\bvec{r}}) Y_{1-1}(\hat{\bvec{a}})
\right.\nonumber\\&&\left.
-\sqrt{\frac{2}{5}}Y_{20}(\hat{\bvec{r}}) Y_{10}(\hat{\bvec{a}})
\right.\nonumber\\&&\left.
+\sqrt{\frac{3}{10}}Y_{2-1}(\hat{\bvec{r}}) Y_{11}(\hat{\bvec{a}})
\right]\nonumber\\
&=&\frac{1}{\sqrt{8\pi}}
\left(x^2 a_z +y^2 a_z - 2z^2 a_z 
\right.\nonumber\\&&\left.
-3yz a_y-3zx a_x \right).
\end{eqnarray}
We define the $x$ component 
\begin{equation}
\bvec{Y}_{1Lx}(\hat{\bvec{r}})\equiv -\frac{1}{2} \left[ 
\bvec{Y}_{1L1}(\hat{\bvec{r}})-\bvec{Y}_{1L-1}(\hat{\bvec{r}})
\right]. 
\end{equation}
and get similar expressions as  
\begin{eqnarray}
\bvec{a}\cdot \left[r^2\bvec{Y}_{10x}(\hat{\bvec{r}})\right]
&=&\frac{1}{\sqrt{4\pi}}
\left(x^2 +y^2+z^2\right) a_x, \\
\bvec{a}\cdot \left[r^2\bvec{Y}_{12x}(\hat{\bvec{r}})\right]
&=&\frac{1}{\sqrt{8\pi}}
\left(y^2 a_x +z^2 a_x - 2x^2 a_x 
\right.\nonumber\\&&\left.
-3zx a_z-3xy a_y \right).
\end{eqnarray}
Here we follow the transformation of the basis set 
$(x,y,z) \to (1,0,-1)$,  
\begin{eqnarray}
\bvec{e}_1&\equiv& -\frac{1}{2}(\bvec{e}_x+i\bvec{e}_y),\nonumber\\
\bvec{e}_0&\equiv& \bvec{e}_z,\nonumber\\
\bvec{e}_{-1}&\equiv& \frac{1}{2}(\bvec{e}_x-i\bvec{e}_y), 
\end{eqnarray}
leading to the relation 
\begin{eqnarray}
a_1&=& -\frac{1}{2}(a_x+ia_y),\nonumber\\
a_0&=& a_z,\nonumber\\
a_{-1}&=& \frac{1}{2}(a_x-ia_y),
\end{eqnarray}
and its inverse relation
\begin{eqnarray}
a_x&=& -\frac{1}{2}(a_1-a_{-1}),\nonumber\\
a_y&=& \frac{i}{2}(a_1+a_{-1}),\nonumber\\
a_z&=& a_0.
\end{eqnarray}

\section{Matrix elements of dipole operators for AMD wave function}

For the single-particle operator of the the current density
\begin{eqnarray}
\bvec{j}_\textrm{sp}(\bvec{r})&\equiv& -\frac{i\hbar}{2m} \nabla_k\delta(\bvec{r}-\bvec{r}_k)+\delta(\bvec{r}-\bvec{r}_k)\nabla_k, 
\end{eqnarray} 
the matrix element for single-particle wave functions 
of the AMD wave function is given as 
\begin{eqnarray}
\langle \varphi_i |\bvec{j}_\textrm{sp}(\bvec{r})|\varphi_j\rangle
&=&\frac{\hbar}{m}\bvec{K}_{ij}
 \phi^*_{\bvec{X}_i}(\bvec{r})  \phi_{\bvec{X}_j}(\bvec{r}) 
\nonumber\\
&\times&
 \langle\chi_i| \chi_j \rangle \langle \tau_i| \tau_j \rangle,\\
\bvec{K}_{ij}&\equiv&-i\sqrt{\nu}(\bvec{X}_i^*-\bvec{X}_j),
\end{eqnarray}
and the matrix element of $ \bvec{j}_\textrm{sp}(\bvec{r})\cdot [r^2 \bvec{Y}_{1 L \mu}(\hat{\bvec{r}})]$
is given as
\begin{eqnarray}
&&\langle \varphi_i | \bvec{j}_\textrm{sp}(\bvec{r})\cdot [r^2 \bvec{Y}_{1 L \mu}(\hat{\bvec{r}})
]|\varphi_j\rangle \nonumber\\
&&\qquad\qquad  = \sqrt{\frac{4\pi}{3}} R^2_{ij}K_{ij}
\left[Y_{L }  ( \hat{\bvec{R}}_{ij}) \otimes Y_{1} ( \hat{\bvec{K}}_{ij})\right] _{1\mu}
\nonumber\\
&&\qquad \qquad \times
 \phi^*_{\bvec{X}_i}(\bvec{r})  \phi_{\bvec{X}_j}(\bvec{r}) \langle \chi_i| \chi_j \rangle \langle \tau_i| \tau_j \rangle,
 \\
&&\bvec{R}_{ij}\equiv\frac{1}{2\sqrt{\nu}}(\bvec{X}_i^*+\bvec{X}_j).
\end{eqnarray}


\begin{thebibliography}{9}

\bibitem{Paar:2007bk} 
  N.~Paar, D.~Vretenar, E.~Khan and G.~Colo,
  %``Exotic modes of excitation in atomic nuclei far from stability,''
  Rept.\ Prog.\ Phys.\  {\bf 70}, 691 (2007).
%  [nucl-th/0701081 [NUCL-TH]].
  %%CITATION = NUCL-TH/0701081;%%
  %202 citations counted in INSPIRE as of 25 Oct 2015

\bibitem{aumann-rev}
T. Aumann and T. Nakamura, 
Phys. Scr. {\bf T152}, 014012 (2013).
%doi:10.1088/0031-8949/2013/T152/014012
%The electric dipole response of
%exotic nuclei

\bibitem{Savran:2013bha} 
  D.~Savran, T.~Aumann and A.~Zilges,
  %``Experimental studies of the Pygmy Dipole Resonance,''
  Prog.\ Part.\ Nucl.\ Phys.\  {\bf 70}, 210 (2013).
%  doi:10.1016/j.ppnp.2013.02.003
  %%CITATION = doi:10.1016/j.ppnp.2013.02.003;%%
  %138 citations counted in INSPIRE as of 13 Apr 2017

\bibitem{Bracco:2015hca} 
  A.~Bracco, F.~C.~L.~Crespi and E.~G.~Lanza,
  %``Gamma decay of pygmy states from inelastic scattering of ions,''
  Eur.\ Phys.\ J.\ A {\bf 51},  99 (2015).
%  doi:10.1140/epja/i2015-15099-6
  %%CITATION = doi:10.1140/epja/i2015-15099-6;%%
  %11 citations counted in INSPIRE as of 13 Apr 2017



\bibitem{Endres:2010zw} 
  J.~Endres {\it et al.},
  %``Isospin Character of the Pygmy Dipole Resonance in 124Sn,''
  Phys.\ Rev.\ Lett.\  {\bf 105}, 212503 (2010)
  doi:10.1103/PhysRevLett.105.212503
  [arXiv:1011.5139 [nucl-ex]].
  %%CITATION = doi:10.1103/PhysRevLett.105.212503;%%
  %95 citations counted in INSPIRE as of 18 Apr 2017

\bibitem{Derya:2014yqk} 
  V.~Derya {\it et al.},
  %``Isospin properties of electric dipole excitations in $^{48}Ca$,''
  Phys.\ Lett.\ B {\bf 730}, 288 (2014)
  doi:10.1016/j.physletb.2014.01.050
  [arXiv:1402.0406 [nucl-ex]].
  %%CITATION = doi:10.1016/j.physletb.2014.01.050;%%
  %21 citations counted in INSPIRE as of 18 Apr 2017

\bibitem{Crespi:2014tka} 
  F.~C.~L.~Crespi {\it et al.},
  %``Isospin Character of Low-Lying Pygmy Dipole States in Pb208 via Inelastic Scattering of O17 Ions,''
  Phys.\ Rev.\ Lett.\  {\bf 113}, no. 1, 012501 (2014).
  doi:10.1103/PhysRevLett.113.012501
  %%CITATION = doi:10.1103/PhysRevLett.113.012501;%%
  %21 citations counted in INSPIRE as of 18 Apr 2017

\bibitem{Crespi:2015sfa} 
  F.~C.~L.~Crespi {\it et al.},
  %``1− and 2+ discrete states in Zr90 populated via the (O17,O′17γ) reaction,''
  Phys.\ Rev.\ C {\bf 91}, no. 2, 024323 (2015).
  doi:10.1103/PhysRevC.91.024323
  %%CITATION = doi:10.1103/PhysRevC.91.024323;%%
  %11 citations counted in INSPIRE as of 18 Apr 2017

\bibitem{Nakatsuka:2017dhs} 
  N.~Nakatsuka {\it et al.},
  %``Observation of isoscalar and isovector dipole excitations in neutron-rich 20 O,''
  Phys.\ Lett.\ B {\bf 768}, 387 (2017).
  doi:10.1016/j.physletb.2017.03.017
  %%CITATION = doi:10.1016/j.physletb.2017.03.017;%%


\bibitem{Berman:1975tt} 
  B.~L.~Berman and S.~C.~Fultz,
  %``Measurements of the giant dipole resonance with monoenergetic photons,''
  Rev.\ Mod.\ Phys.\  {\bf 47}, 713 (1975).
  %%CITATION = RMPHA,47,713;%%
  %274 citations counted in INSPIRE as of 24 Oct 2015

\bibitem{morsch80}
H.P. Morsch, M. Rogge, P. Turek, and C. Mayer-B\''oricke, Phys. Rev. Lett. {\bf 45}, 337 (1980).

\bibitem{Harakeh:1981zz} 
  M.~N.~Harakeh and A.~E.~L.~Dieperink,
  %``Isoscalar dipole resonance: Form factor and energy weighted sum rule,''
  Phys.\ Rev.\ C {\bf 23}, 2329 (1981).
 % doi:10.1103/PhysRevC.23.2329
  %%CITATION = doi:10.1103/PhysRevC.23.2329;%%
  %68 citations counted in INSPIRE as of 06 Dec 2015

\bibitem{giai81}
%``MONOPOLE AND DIPOLE COMPRESSION MODES IN NUCLEI
N. V. Giai and H. Sagawa,
Nucl. Phys. {\bf A371}, 1 (1981).


\bibitem{Colo:2003zm} 
  G.~Colo and G.~N.~Van,
  %``Theoretical understanding of the nuclear incompressibility: Where do we stand?,''
  Nucl.\ Phys.\ A {\bf 731}, 15 (2004).
%  doi:10.1016/j.nuclphysa.2003.11.014
%  [nucl-th/0309002].
  %%CITATION = doi:10.1016/j.nuclphysa.2003.11.014;%%
  %35 citations counted in INSPIRE as of 06 Dec 2015

\bibitem{Decowski:1981pcz} 
  P.~Decowski, H.~P.~Morsch and W.~Benenson,
  %``Low-lying isoscalar dipole excitations in 208 Pb,''
  Phys.\ Lett.\  {\bf 101B}, 147 (1981).
%  doi:10.1016/0370-2693(81)90661-4
  %%CITATION = doi:10.1016/0370-2693(81)90661-4;%%
  %7 citations counted in INSPIRE as of 17 Apr 2017

\bibitem{Poelhekken:1992gvp} 
  T.~D.~Poelhekken, S.~K.~B.~Hesmondhalgh, H.~J.~Hofmann, A.~van der Woude and M.~N.~Harakeh,
  %``Low-energy isoscalar dipole strength in 40 Ca, 58 Ni, 90 Zr and 208 Pb,''
  Phys.\ Lett.\ B {\bf 278}, 423 (1992).
%  doi:10.1016/0370-2693(92)90579-S
  %%CITATION = doi:10.1016/0370-2693(92)90579-S;%%
  %54 citations counted in INSPIRE as of 17 Apr 2017

\bibitem{Colo:2000br} 
  G.~Colo, N.~Van Giai, P.~F.~Bortignon and M.~R.~Quaglia,
  %``On dipole compression modes in nuclei,''
  Phys.\ Lett.\ B {\bf 485}, 362 (2000).
%  doi:10.1016/S0370-2693(00)00725-5
% [nucl-th/0003042].
  %%CITATION = doi:10.1016/S0370-2693(00)00725-5;%%
  %81 citations counted in INSPIRE as of 14 Apr 2017

\bibitem{Vretenar:2001te} 
  D.~Vretenar, N.~Paar, P.~Ring, and T. Nik\u{s}i\'{c}
  %``Toroidal dipole resonances in the relativistic random phase approximation,''
  Phys.\ Rev.\ C {\bf 65}, 021301 (2002).
%  doi:10.1103/PhysRevC.65.021301
%  [nucl-th/0107024].
  %%CITATION = doi:10.1103/PhysRevC.65.021301;%%
  %53 citations counted in INSPIRE as of 14 Apr 2017

\bibitem{Ryezayeva:2002zz} 
  N.~Ryezayeva {\it et al.},
  %``Nature of Low-Energy Dipole Strength in Nuclei: The Case of a Resonance at Particle Threshold in P-208b,''
  Phys.\ Rev.\ Lett.\  {\bf 89}, 272502 (2002).
  %%CITATION = PRLTA,89,272502;%%
  %123 citations counted in INSPIRE as of 25 Oct 2015


\bibitem{Papakonstantinou:2010ja} 
  P.~Papakonstantinou, V.~Y.~Ponomarev, R.~Roth and J.~Wambach,
  %``Isoscalar dipole coherence at low energies and forbidden E1 strength,''
  Eur.\ Phys.\ J.\ A {\bf 47}, 14 (2011).
%  doi:10.1140/epja/i2011-11014-7
%  [arXiv:1011.1162 [nucl-th]].
  %%CITATION = doi:10.1140/epja/i2011-11014-7;%%
  %10 citations counted in INSPIRE as of 14 Apr 2017



\bibitem{Kvasil:2011yk} 
  J.~Kvasil, V.~O.~Nesterenko, W.~Kleinig, P.-G.~Reinhard and P.~Vesely,
  %``General Treatment of Vortical, Toroidal, and Compression Modes,''
  Phys.\ Rev.\ C {\bf 84}, 034303 (2011).
%  doi:10.1103/PhysRevC.84.034303
%  [arXiv:1105.0837 [nucl-th]].
  %%CITATION = doi:10.1103/PhysRevC.84.034303;%%
  %14 citations counted in INSPIRE as of 10 Apr 2017

\bibitem{Repko:2012rj} 
  A.~Repko, P.-G.~Reinhard, V.~O.~Nesterenko and J.~Kvasil,
  %``Toroidal nature of the low-energy $E1$ mode,''
  Phys.\ Rev.\ C {\bf 87},  024305 (2013).
%  doi:10.1103/PhysRevC.87.024305
%  [arXiv:1212.2088 [nucl-th]].
  %%CITATION = doi:10.1103/PhysRevC.87.024305;%%
  %26 citations counted in INSPIRE as of 13 Apr 2017

\bibitem{Nesterenko:2016qiw} 
  V.~O.~Nesterenko, J.~Kvasil, A.~Repko, W.~Kleinig and P.-G.~Reinhard,
  %``Toroidal resonance: relation to pygmy mode, vortical properties and anomalous deformation splitting,''
  Phys.\ Atom.\ Nucl.\  {\bf 79}, 842 (2016).
%  doi:10.1134/S106377881606020X
%  [arXiv:1602.03326 [nucl-th]].
  %%CITATION = doi:10.1134/S106377881606020X;%%



\bibitem{Suzuki:1989zza} 
  Y.~Suzuki and S.~Hara,
  %``Isoscalar monopole and quadrupole strength of O-16 in an alpha+ C-12 cluster and symplectic mixed basis,''
  Phys.\ Rev.\ C {\bf 39}, 658 (1989).
 % doi:10.1103/PhysRevC.39.658
  %%CITATION = doi:10.1103/PhysRevC.39.658;%%
  %6 citations counted in INSPIRE as of 15 feb. 2016
\bibitem{Kawabata:2005ta} 
  T.~Kawabata {\it et al.},
  %``Indication of dilute 2 alpha + t cluster structure in B-11,''
  Phys.\ Lett.\ B {\bf 646}, 6 (2007).
%  doi:10.1016/j.physletb.2006.11.079
%  [nucl-ex/0512040].
  %%CITATION = doi:10.1016/j.physletb.2006.11.079;%%
  %42 citations counted in INSPIRE as of 15 Feb 2016

\bibitem{Funaki:2006gt} 
  Y.~Funaki, A.~Tohsaki, H.~Horiuchi, P.~Schuck and G.~Ropke,
  %``Inelastic form-factors to alpha particle condensate states in C-12 and O-16: What can we learn?,''
  Eur.\ Phys.\ J.\ A {\bf 28}, 259 (2006).
%  doi:10.1140/epja/i2006-10061-5
 % [nucl-th/0601035].
  %%CITATION = doi:10.1140/epja/i2006-10061-5;%%
  %21 citations counted in INSPIRE as of 06 Dec 2015


\bibitem{KanadaEn'yo:2006bd} 
  Y.~Kanada-En'yo,
  %``Negative parity states of B-11 and C-11 and the similarity with C-12,''
  Phys.\ Rev.\ C {\bf 75}, 024302 (2007).
%  [nucl-th/0612109].
  %%CITATION = NUCL-TH/0612109;%%
  %17 citations counted in INSPIRE as of 18 Jul 2013

\bibitem{Wakasa:2007zza} 
  T.~Wakasa, E.~Ihara, K.~Fujita, Y.~Funaki, K.~Hatanaka, H.~Horiuchi, M.~Itoh and J.~Kamiya {\it et al.},
  %``New candidate for an alpha cluster condensed state in O-16(alpha, alpha') at 400-MeV,''
  Phys.\ Lett.\ B {\bf 653}, 173 (2007).
  %%CITATION = PHLTA,B653,173;%%
  %19 citations counted in INSPIRE as of 18 Jul 2013



\bibitem{Yamada:2011ri} 
  T.~Yamada, Y.~Funaki, T.~Myo, H.~Horiuchi, K.~Ikeda, G.~Ropke, P.~Schuck and A.~Tohsaki,
  %``Isoscalar monopole excitations in $^{16}$O: $\alpha$-cluster states at low energy and mean-field-type states at higher energy,''
  Phys.\ Rev.\ C {\bf 85}, 034315 (2012).
%  doi:10.1103/PhysRevC.85.034315
 % [arXiv:1110.6509 [nucl-th]].
  %%CITATION = doi:10.1103/PhysRevC.85.034315;%%
  %16 citations counted in INSPIRE as of 06 Dec 2015


\bibitem{Ichikawa:2011ji} 
  T.~Ichikawa, N.~Itagaki, T.~Kawabata, T.~Kokalova and W.~von Oertzen,
  %``Gas-like state of $\alpha$ clusters around $^{16}$O core in $^{24}$Mg,''
  Phys.\ Rev.\ C {\bf 83}, 061301 (2011).
%  doi:10.1103/PhysRevC.83.061301
%  [arXiv:1102.1477 [nucl-th]].
  %%CITATION = doi:10.1103/PhysRevC.83.061301;%%
  %7 citations counted in INSPIRE as of 06 Dec 2015 
\bibitem{kawabata13}
T. Kawabata {\it et al.},  Jour. Phys. Conf. Ser. {\bf 436}, 012009 (2013).

\bibitem{Kanada-En'yo:2013dma} 
  Y.~Kanada-En'yo,
  %``Cluster states and monopole transitions in $^{16}$O,''
  Phys.\ Rev.\ C {\bf 89}, 024302 (2014).
  %[arXiv:1308.0392 [nucl-th]].
  %%CITATION = ARXIV:1308.0392;%%
  %8 citations counted in INSPIRE as of 25 Oct 2015



\bibitem{Kanada-Enyo:2015vwc} 
  Y.~Kanada-En'yo,
  %``Isoscalar monopole and dipole excitations of cluster states and giant resonances in $^{12}$C,''
  Phys.\ Rev.\ C {\bf 93}, 054307 (2016).
%  doi:10.1103/PhysRevC.93.054307
%  [arXiv:1512.03619 [nucl-th]].
  %%CITATION = doi:10.1103/PhysRevC.93.054307;%%
  %6 citations counted in INSPIRE as of 14 Apr 2017



\bibitem{Chiba:2015zxa} 
  Y.~Chiba and M.~Kimura,
  %``Cluster states and isoscalar monopole transitions of $^{24}$Mg,''
  Phys.\ Rev.\ C {\bf 91}, 061302 (2015).
%  doi:10.1103/PhysRevC.91.061302
%  [arXiv:1502.06325 [nucl-th]].
  %%CITATION = doi:10.1103/PhysRevC.91.061302;%%

\bibitem{Chiba:2015khu} 
  Y.~Chiba, M.~Kimura and Y.~Taniguchi,
  %``Isoscalar dipole transition as a probe for asymmetric clustering,''
  Phys.\ Rev.\ C {\bf 93}, 034319 (2016).
%  doi:10.1103/PhysRevC.93.034319
%  [arXiv:1512.08214 [nucl-th]].
  %%CITATION = doi:10.1103/PhysRevC.93.034319;%%
  %7 citations counted in INSPIRE as of 14 Apr 2017

\bibitem{Chiba:2016zyz} 
  Y.~Chiba, Y.~Taniguchi and M.~Kimura,
  %``Inversion doublets of reflection-asymmetric clustering in 28Si and their isoscalar monopole and dipole transitions,''
  arXiv:1610.04000 [nucl-th].
  %%CITATION = ARXIV:1610.04000;%%
  %1 citations counted in INSPIRE as of 14 Apr 2017



\bibitem{Kimura:2016heo} 
  M.~Kimura,
  %``Structure and decay of the pygmy dipole resonance in 26Ne,''
  Phys.\ Rev.\ C {\bf 95}, no. 3, 034331 (2017).
%  doi:10.1103/PhysRevC.95.034331
%  [arXiv:1611.08804 [nucl-th]].
  %%CITATION = doi:10.1103/PhysRevC.95.034331;%%

\bibitem{John:2003ke} 
  B.~John, Y.~Tokimoto, Y.-W.~Lui, H.~L.~Clark, X.~Chen and D.~H.~Youngblood,
  %``Isoscalar electric multipole strength in C-12,''
  Phys.\ Rev.\ C {\bf 68}, 014305 (2003).
 % doi:10.1103/PhysRevC.68.014305
  %%CITATION = doi:10.1103/PhysRevC.68.014305;%%
  %31 citations counted in INSPIRE as of 30 Nov 2015


\bibitem{Lui:2001xh} 
  Y.-W.~Lui, H.~L.~Clark and D.~H.~Youngblood,
  %``Giant resonances in 16o,''
  Phys.\ Rev.\ C {\bf 64}, 064308 (2001).
%  doi:10.1103/PhysRevC.64.064308
  %%CITATION = doi:10.1103/PhysRevC.64.064308;%%
  %29 citations counted in INSPIRE as of 06 Dec 2015




\bibitem{Oertzen-rev}
W. von Oertzen, M. Freer and Y. Kanada-En'yo, Phys. Rep. {\bf 432}, 43 (2006).

\bibitem{freer07-rev}
M. Freer, 
Rep. Prog. Phys. {\bf 70}, 2149 (2007).
% doi:10.1088/0034-4885/70/12/R03
%The clustered nucleus—cluster structures in stable
%and unstable nuclei

\bibitem{KanadaEn'yo:2012bj}
  Y.~Kanada-En'yo, M.~Kimura and A.~Ono,
  %``Antisymmetrized molecular dynamics and its applications to cluster phenomena,''
  PTEP {\bf 2012},  01A202 (2012).
%  [arXiv:1202.1864 [nucl-th]].
  %%CITATION = ARXIV:1202.1864;%%
  %1 citations counted in INSPIRE as of 28 Jul 2013



\bibitem{Ito2014-rev}
M. Ito and K. Ikeda, 
Rep. Prog. Phys. {\bf 77}, 096301 (2014) .
%Unified studies of chemical bonding structures and resonant scattering in light neutron-excess
%systems, 10,12Be


\bibitem{Ito:2011zza} 
  M.~Ito,
  %``Imprint of adiabatic structures in monopole excitations of Be-12,''
  Phys.\ Rev.\ C {\bf 83}, 044319 (2011).
%  doi:10.1103/PhysRevC.83.044319
  %%CITATION = doi:10.1103/PhysRevC.83.044319;%%
  %12 citations counted in INSPIRE as of 11 fev. 2016


\bibitem{Kanada-Enyo:2016jnq} 
  Y.~Kanada-En'yo,
  %``Monopole transitions to cluster states in $^{10}$Be and $^9$Li,''
  Phys.\ Rev.\ C {\bf 94},  024326 (2016).
%  doi:10.1103/PhysRevC.94.024326
% [arXiv:1604.01453 [nucl-th]].
  %%CITATION = doi:10.1103/PhysRevC.94.024326;%%
  %1 citations counted in INSPIRE as of 14 Apr 2017

\bibitem{Kanada-Enyo:2015knx} 
  Y.~Kanada-En'yo,
  %``Isovector and isoscalar dipole excitations in $^{9}$Be and $^{10}$Be studied with antisymmetrized molecular dynamics,''
  Phys.\ Rev.\ C {\bf 93}, 024322 (2016).
%  doi:10.1103/PhysRevC.93.024322
% [arXiv:1511.08530 [nucl-th]].
  %%CITATION = doi:10.1103/PhysRevC.93.024322;%%
  %3 citations counted in INSPIRE as of 09 Apr 2017file:///C:/Users/kanad/Dropbox/isd-tm/be10/paper/be10-dipole.tex
%\input references.tex

\bibitem{Ono:1991uz} 
  A.~Ono, H.~Horiuchi, T.~Maruyama and A.~Ohnishi,
  %``Fragment formation studied with antisymmetrized version of molecular dynamics with two nucleon collisions,''
  Phys.\ Rev.\ Lett.\  {\bf 68}, 2898 (1992).
  %%CITATION = PRLTA,68,2898;%%
  %121 citations counted in INSPIRE as of 25 Oct 2015

\bibitem{Ono:1992uy} 
  A.~Ono, H.~Horiuchi, T.~Maruyama and A.~Ohnishi,
  %``Antisymmetrized version of molecular dynamics with two nucleon collisions and its application to heavy ion reactions,''
  Prog.\ Theor.\ Phys.\  {\bf 87}, 1185 (1992).
  %%CITATION = PTPKA,87,1185;%%
  %125 citations counted in INSPIRE as of 25 Oct 2015

\bibitem{KanadaEnyo:1995tb}
  Y.~Kanada-En'yo, H.~Horiuchi and A.~Ono,
  %``Structure of Li and Be isotopes studied with antisymmetrized molecular
  %dynamics,''
  Phys.\ Rev.\  C {\bf 52}, 628  (1995).
  %%CITATION = PHRVA,C52,628;%%

\bibitem{KanadaEnyo:1995ir}
  Y.~Kanada-En'yo and H.~Horiuchi,
%  %``Neutron-Rich B Isotopes Studied With Antisymmetrized Molecular Dynamics,''
  Phys.\ Rev.\  C {\bf 52}, 647 (1995).
%  %%CITATION = PHRVA,C52,647;%%


%\bibitem{AMDsupp} 
\bibitem{KanadaEn'yo:2001qw} 
  Y.~Kanada-En'yo and H.~Horiuchi,
  %``Structure of light unstable nuclei studied with antisymmetrized molecular dynamics,''
  Prog.\ Theor.\ Phys.\ Suppl.\  {\bf 142}, 205 (2001).
%  doi:10.1143/PTPS.142.205
%  [nucl-th/0107044].
  %%CITATION = doi:10.1143/PTPS.142.205;%%
  %93 citations counted in INSPIRE as of 13 feb. 2016



\bibitem{Dubovik75}
V.M. Dubovik and A.A. Cheshkov, Sov. J. Part. Nucl. {\bf 5},
318 (1975).

\bibitem{semenko81}
S. F. Semenko,  Sov. J. Nucl. Phys. {\bf 34}, 356 (1981).


\bibitem{Ravenhall:1987thb} 
  D.~G.~Ravenhall and J.~Wambach,
  %``Nuclear transition currents and vorticity,''
  Nucl.\ Phys.\ A {\bf 475}, 468 (1987).
%  doi:10.1016/0375-9474(87)90074-1
  %%CITATION = doi:10.1016/0375-9474(87)90074-1;%%
  %17 citations counted in INSPIRE as of 14 Apr 2017

\bibitem{TOHSAKI}
 T. Ando, K.Ikeda, and A. Tohsaki, Prog. Theor. Phys.
 {\bf 64}, 1608 (1980).
\bibitem{LS1}
 R. Tamagaki, Prog. Theor. Phys. {\bf 39}, 91 (1968).

\bibitem{LS2}
 N. Yamaguchi, T. Kasahara, S. Nagata, and Y. Akaishi,
 Prog. Theor. Phys. {\bf 62}, 1018 (1979).


\bibitem{KanadaEn'yo:1998rf} 
  Y.~Kanada-En'yo,
  %``Variation after angular momentum projection for the study of excited states based on antisymmetrized molecular dynamics,''
  Phys.\ Rev.\ Lett.\  {\bf 81}, 5291 (1998).
%  [nucl-th/0204039].
  %%CITATION = NUCL-TH/0204039;%%
  %74 citations counted in INSPIRE as of 01 Nov 2014  

\bibitem{KanadaEn'yo:1999ub} 
  Y.~Kanada-En'yo, H.~Horiuchi and A.~Dote,
  %``Structure of excited states of Be-10 studied with antisymmetrized molecular dynamics,''
  Phys.\ Rev.\ C {\bf 60}, 064304 (1999).
  %[nucl-th/9905048].
  %%CITATION = NUCL-TH/9905048;%%
  %83 citations counted in INSPIRE as of 25 Oct 2015
\bibitem{KanadaEn'yo:2006ze} 
  Y.~Kanada-En'yo,
  %``Structure of ground and excited states of C-12,''
  Prog.\ Theor.\ Phys.\  {\bf 117}, 655 (2007)
  [Prog.\ Theor.\ Phys.\  {\bf 121}, 895 (2009)].
  %[nucl-th/0605047].
  %%CITATION = NUCL-TH/0605047;%%
  %74 citations counted in INSPIRE as of 25 Oct 2015




%\bibitem{Kvasil:2012kc} 
%  J.~Kvasil, V.~O.~Nesterenko, A.~Repko, W.~Kleinig, P.-G.~Reinhard and N.~L.~Iudice,
  %``Toroidal, compression, and vortical dipole strengths in 124Sn,''
%  Phys.\ Scripta T {\bf 154}, 014019 (2013)
%  doi:10.1088/0031-8949/2013/T154/014019
%  [arXiv:1211.2978 [nucl-th]].
  %%CITATION = doi:10.1088/0031-8949/2013/T154/014019;%%
  %7 citations counted in INSPIRE as of 13 Apr 2017





%\include{references}

\end{thebibliography}
\end{document}